\documentclass[prl,twocolumn,showpacs,preprintnumbers,amsmath,amssymb]{revtex4}
\usepackage{graphicx} 

\usepackage[T1]{fontenc}
\usepackage[ansinew]{inputenc} 

\usepackage{textcomp}
\usepackage{gensymb} 

\usepackage{lipsum,mathptmx,etoolbox}

\usepackage{calc}
\usepackage{graphicx}
\usepackage{amsmath}
\usepackage{amssymb}
\usepackage{gensymb}
\usepackage{epsfig}
\usepackage{amsthm}
\usepackage{bm}
\usepackage{color}
\usepackage[toc,page]{appendix}
\usepackage{ragged2e}

\usepackage{overpic}

\usepackage{xr}

\def\be{\begin{equation}}	
\def\ee{\end{equation}}
\def\arr{\begin{array}{rll}}
\def\ea{\end{array}}
\def\bea{\begin{eqnarray}}
\def\eea{\end{eqnarray}}

\begin{document}

\title{Nanoroughness induced anti-reflection and haze effects in opaque systems}
\author{V.Gareyan$^{1}$, N.Margaryan$^{1}$ and Zh.Gevorkian$^{1,2}$}
\affiliation{$^{1}$ Alikhanyan National Laboratory, Alikhanian Brothers St. 2,  Yerevan 0036 Armenia\\
$^{2}$ Institute of Radiophysics and Electronics, Ashtarak-2 0203 Armenia}
\begin{abstract}
How to make a material anti-reflective without changing its high refraction index? Achieving anti-reflection in high-refractive-index materials poses challenges due to their high reflectivity (Fresnel equations). Based on theory with new boundary conditions, we propose modifying surface properties on a nanoscale to tackle this. Our study on weakly rough opaque surfaces reveals significant changes in specular and diffuse scattering, predicting anti-reflection where roughness matches light penetration depth for the first time. Experimental validation on nano-roughened \textit{Si} films (at wavelengths 300-400 nm) supports our findings. We also analyze angular and polarization dependences of nanoroughness-induced haze, showing predominant p-polarization and minimal haze at nanoscale, yet impactful specular reflection reduction.
\end{abstract}

\maketitle

\textit{Introduction}-The antireflection effect has garnered significant attention in modern technologies \cite{rad2020,hossain2021,isakov2020,wang2024,chen2001,ji2022,chen2000,mehmood2016,chen2011,kim2013}, prompting extensive research efforts aimed at developing a comprehensive theory capable of explaining vast experimental data \cite{amra1986,poitras2004,moys1974,kozhevnikov2024,ong2024}. Often, materials requiring antireflective properties possess high refractive indices, posing challenges as per Fresnel formulas. This paper addresses these challenges by investigating corrections to reflection coefficients induced by surface roughness. Incident light on rough surfaces generates specular and diffusive scattered waves, detailed in \cite{simonsen2010,hormann2023,lamare2023,ciochon2023,hanson2024,ballington2024}. When incident light wavelengths greatly exceed surface roughness characteristics, diffuse components diminish, favoring predominantly specular reflection. This study focuses on reflection from weakly rough opaque systems, which, as recently highlighted \cite{gevorkian2022}, exhibit distinct behaviors compared to transparent systems. Specifically, we demonstrate how nanoroughness in opaque systems significantly reduces specular reflection, keeping the haze small. The concept of haze or "cloudiness" is introduced to quantify the portion of diffusively scattered light provided the weak roughness of surface \cite{simonsen2009,astm2000,billmeyer1985}.

We particularly examine silicon, a material crucial in micro- and nanoelectronics, emphasizing the importance of accurately diagnosing substrate surface roughness for ensuring product quality \cite{mori2020,wang2005,khan2007}. Our attention is drawn to nano-roughness scenarios, where surface profile root mean square (RMS) values are in the nanometer range. For systems like air-bulk silicon with substantial dielectric constant discrepancies, precise boundary adjustments are essential for achieving accurate results \cite{carminati2020}. This disparity in refractive indices between media plays a pivotal role in achieving effective antireflective properties.

Various nanotexturing techniques, such as plasma etching \cite{moreno2010,armstrong2018} and acid-based wet etching \cite{stepf2017,zrir2024,dimitrov2013}, are employed to control surface properties. Diffuse reflection spectroscopy is instrumental in characterizing the optical behaviors of rough surfaces \cite{moreno2010,zrir2024,dimitrov2013}. By assessing reflection changes before and after etching, the impact of surface roughness on light scattering and absorption can be analyzed. Detailed surface morphology studies are conducted using techniques such as scanning electron microscopy (SEM) \cite{moreno2010,stepf2017,zrir2024,dimitrov2013}, Confocal Scanning Microscopy (CLSM) \cite{stepf2017}, and Atomic Force Microscopy (AFM) \cite{armstrong2018,pasanen2020,islam2015}. The combined use of diffuse reflection spectroscopy and AFM provides comprehensive insights into surface roughness characteristics.

In this study, we initially assume Gaussian roughness, corroborated by AFM data analysis, demonstrating its appropriateness and justification.

\begin{figure}
    \centering
    \includegraphics[width=0.9\columnwidth]{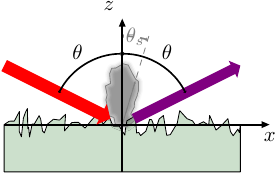}
    \caption{This schematic illustrates light incidence on a rough surface, where \(\theta\) denotes the angle of incidence of the light beam. The incident beam is shown in red, the specular reflection is depicted in violet, and the diffuse part (haze) is represented by a gray cloud. The haze intensity is plotted as a polar plot along some direction, characterized by the polar angle \(\theta_s\). The azimuthal angle \(\phi_s\) is not shown for clarity.}
    \label{illustration}
\end{figure}

\textit{Theory}-We start from the perturbative solution of the wave equation with modified boundaries (see in \cite{carminati2020}):

\begin{eqnarray}
    \nabla^2 \textbf{E} - \nabla (\nabla \cdot \textbf{E}) + k_0^2 \varepsilon(\boldsymbol{\rho},z) \textbf{E} = 0,
\end{eqnarray}

where the inhomogeneous \color{black} dielectric constant  $\varepsilon(\boldsymbol{\rho},z)=\varepsilon_0(\boldsymbol{\rho},z)+\varepsilon_1(\boldsymbol{\rho},z)$ and 

\begin{align}
\varepsilon_0(\boldsymbol{\rho},z)=\left\{\begin{array}{cc}
         1, \mbox{~~~~for~} z>h(\boldsymbol{\rho}) \\
         \varepsilon, \mbox{~~~~for~} z<h(\boldsymbol{\rho})
         \end{array}\right.
\label{epszero}
\end{align}

together with

\begin{equation}
\varepsilon_1(\boldsymbol{\rho},z)=[\varepsilon-1]\times \delta(z-h(\boldsymbol{\rho})) \times h(\boldsymbol{\rho}),
\label{eps1}
\end{equation}

where we denote $\varepsilon(\omega) \equiv \varepsilon$ for brevity (see  in \cite{gevorkian2024}). $h(\boldsymbol{\rho})$ is a random function, which describes the rough surface and the respective correlation function is assumed to be Gaussian. The reference field in the case of a P-polarized wave:

\begin{align}
E_{0p,x}({\bf r})=\left\{\begin{array}{c}\cos\theta \times (e^{ik_xx-ik_{z}z}-r_p e^{ik_xx+ik_{z}z}),\\ \mbox{~for~} z>h(x,y),
\\ \\
t_p\cos\theta_t \times e^{ik_xx+ik_{-}z}, \\ \mbox{\qquad\qquad for~} z<h(x,y),
\end{array}\right.
\label{backx}
\end{align}
and
\begin{align}
E_{0p,z}({\bf r})=\left\{\begin{array}{c}\sin\theta \times(e^{ik_xx-ik_{z}z}+r_p e^{ik_xx+ik_{z}z}),\\ \mbox{~for~} z>h(x,y),
\\ \\
t_p \sin\theta_t \times e^{ik_xx+ik_{-}z}, \\ \mbox{\qquad\qquad for~} z<h(x,y),
\end{array}\right.
\label{backz}
\end{align}
and $\sin\theta_t=\sin\theta/\sqrt{\varepsilon}$, $ \cos\theta_t=\sqrt{1-\sin^2\theta/\varepsilon}$, $\theta_t$ is the transmission angle and is generally complex and $\break k_{-}=-k_0 \sqrt{\varepsilon-\sin^2\theta}$, $k_x=k_0 \sin\theta$, $k_z=k_0 \cos\theta$, $k_0 = \omega/c$ and $r_p$ and $t_p$ are the reflection and transmission coefficients from a smooth surface for a p-polarized wave:

\begin{eqnarray}
 r_p=\frac{\varepsilon \cos\theta- \sqrt{\varepsilon-\sin^2\theta}}{\varepsilon \cos\theta+\sqrt{\varepsilon-\sin^2\theta}}, \nonumber\\
 t_p=\frac{2\sqrt{\varepsilon}\cos\theta}{\varepsilon \cos\theta+\sqrt{\varepsilon-\sin^2\theta}}.
\label{ppolar}
\end{eqnarray}

In previous works, the shift of the reference fields to the actual boundary was not applied, which led to high discrepancy compared with the experimental results. Currently, the modified boundary conditions for the reference fields lead to a drastic difference in results compared to other theories \cite{navarrete2009}.
\color{black}

We solve the system using the perturbative approach, where the total field is decomposed with respect to the powers of $h(\boldsymbol{\rho})$. After averaging over different possible realizations of the rough surface only even powers will survive. We will keep only the terms up to $\sim \delta^2$. This allows us to simplify calculations, and in particular, the decomposition of the reflection coefficients into the specular and diffuse parts becomes equivalent to the decomposition into the coherent and incoherent parts. For more details see Supplemental Material.

The final expression for the cases of P- and S-polarized incidence are ${\bf R}_{\alpha}^{t}=\left<{\bf R}_{\alpha}^{spec}\right>+\left<{\bf R}_{p\alpha}^{diff}\right>+\left<{\bf R}_{s\alpha}^{diff}\right>$, where


\begin{widetext}
\begin{gather}
    \left<{\bf R}_{p}^{spec}\right>=|r_p|^2 + 2\left(\frac{\delta}{a}\right)^2 Im\left[ r_p^* t_p\frac{(\varepsilon-1)^2}{\cos\theta} (G_{xx}^{(+-)}, 0, G_{xz}^{(+-)})|_{r=\beta\sin\theta} \cdot \hat{I}^{(2D)}(\beta,\theta)\cdot \left(\begin{array}{c}
        \sqrt{1-\sin^2\theta / \varepsilon} \\
        0 \\
        \sin\theta / \sqrt{\varepsilon}
    \end{array}\right) \right]-\nonumber\\
    -4k_0\frac{\delta^2}{a} Re\left[ r_p^* t_p\frac{(\varepsilon-1)}{\cos\theta} \sqrt{\varepsilon-\sin^2\theta} \left( G_{xx}^{(+-)}(r=\beta\sin\theta)\sqrt{1-\frac{\sin^2\theta}{\varepsilon}} + G_{xz}^{(+-)}(r=\beta\sin\theta) \frac{\sin\theta}{\sqrt{\varepsilon}} \right) \right],\nonumber\\
    \left<{\bf R}_{s}^{spec}\right>=|r_s|^2-4k_0\frac{\delta^2}{a}Re\left[r_s^* t_s (\varepsilon-1)\sqrt{\varepsilon-\sin^2\theta}G_{yy}^{(+-)}(r=\beta\sin\theta)\right]+2\left(\frac{\delta}{a}\right)^2Im\left[ r_s^* t_s G_{yy}^{(+-)}(r=\beta\sin\theta) I^{(2D)}_{yy}(\beta,\theta) \right],
    \label{RPandStotal}
\end{gather}
\end{widetext}

where $\beta=k_0 a$, are the specular reflection coefficients and the expressions for components of tensors $\hat{G}^{(+-)}$ and $\hat{I}^{(2D)}$ as well as the expressions for the diffuse coefficients are not given here for brevity. They can be found in the Supplemental Material. Note, that all figures were plotted using the exact formulas.\color{black}

\begin{figure}
\begin{tabular}{c}
    \begin{overpic}[width=0.9\columnwidth,keepaspectratio]{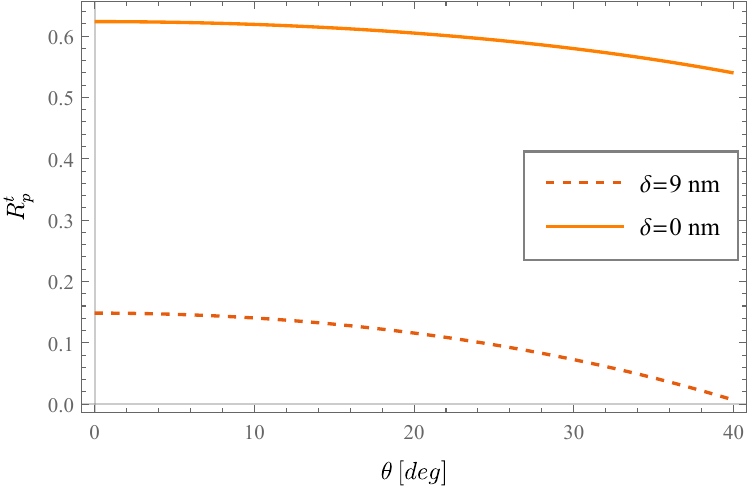}
    \put(0,0.56\columnwidth){(a)}
\end{overpic} \\
    \begin{overpic}[width=0.9\columnwidth,keepaspectratio]{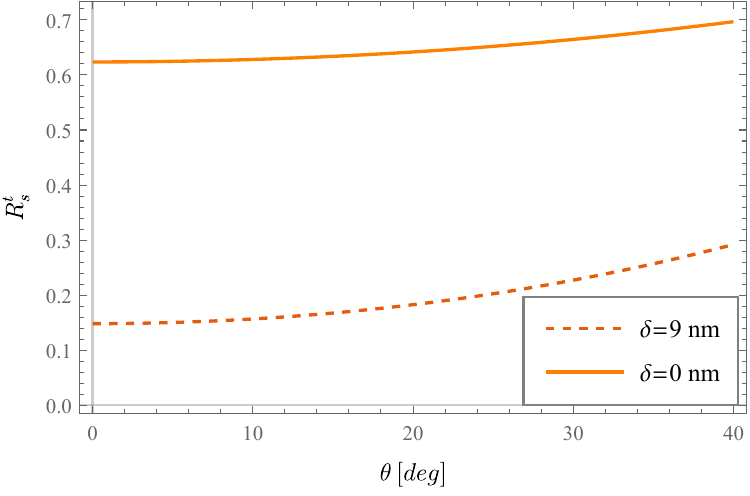}
    \put(0,0.56\columnwidth){(b)}
\end{overpic}\\
    \begin{overpic}[width=0.9\columnwidth,keepaspectratio]{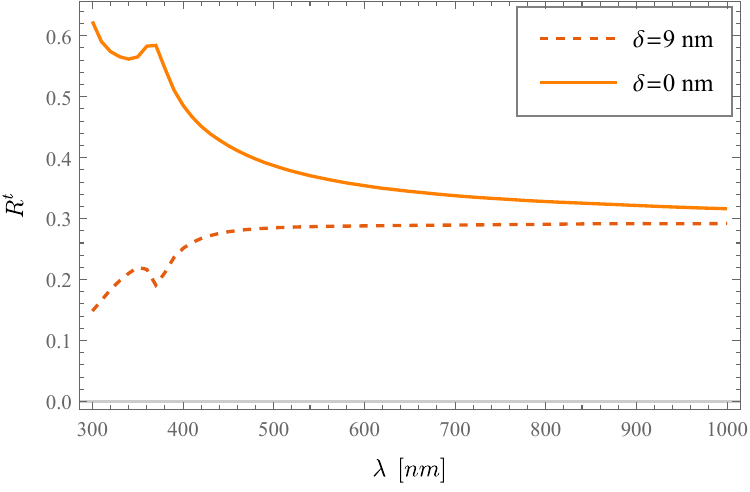}
    \put(0,0.56\columnwidth){(c)}
\end{overpic}\\
\end{tabular}
\caption{The numerically obtained total reflection coefficients are shown for P-polarized (a) and S-polarized (b) incidence of a wave with a wavelength of 300 nm on a silicon surface. The plot distinguishes between reflections from a flat surface (continuous line) and a rough surface (dashed line), varying with the angle of oblique incidence \(\theta\). Additionally, (c) displays the total reflection coefficient for normal incidence as a function of the wavelength \(\lambda\). The correlation length used in the simulations is 50 nm.}
\label{RoughVsFlat}
\end{figure}

In case of a normal incidence and $\beta \gg 1$ these formulas reduce to:

\begin{gather}
    \left<{\bf R}^{spec}\right>=\left|\frac{\sqrt{\varepsilon}-1}{\sqrt{\varepsilon}+1}\right|^2 [1-4k_0^2\delta^2 (n+1)],\nonumber\\
    \left<{\bf R}^{diff}\right>=2k_0^2\delta^2\left|1-\frac{1}{\sqrt{\varepsilon}}\right|^2
    \label{ApproxBigBeta}
\end{gather}

where $n=Re(\sqrt{\varepsilon})$ is the refraction index. This result is a good approximation for the graphs in Fig. \ref{RoughVsFlat}, although they were numerically calculated for $\beta \sim 1$, which corresponds to the values in our experiment. Because of the multiplier $n$, which does not appear in the theory by Navarrete \textit{et al.} (2009) \cite{navarrete2009}, the reflection reduces significantly. Following Eq.(\ref{ApproxBigBeta}) one could think that it is possible to easily reduce reflection by increasing the roughness RMS. However, the latter leads to the respective increase of the diffusive scattering. Our point is that there is an optimal roughness where the specular reflection is reduced significantly with the haze being kept small enough, because it does not contain the large multiplier $n$, see Eq. (\ref{ApproxBigBeta}) and Eq. (\ref{estimate}). 

\textit{Experiment}-The 10 \(\mathrm{\mu}\)m thick monocrystalline silicon wafers with well-polished surfaces (University Wafers, ID
3728, orientation <100>) were first cleaned using a plasma cleaning method to remove any
surface contaminants. The etching solution was prepared by mixing 4.57 vol$\%$ hydrofluoric acid
(HF, 49$\%$), 4.57 vol$\%$ hydrogen peroxide ($\text{H}_2\text{O}_2$, 30$\%$), and 49.10 vol$\%$ hydrochloric acid (HCl,
37$\%$) in deionized water. The silicon wafers were immersed in this mixture for 1 second at room
temperature to achieve a roughness of 8-9 nm. After etching, the wafers were immediately
rinsed with deionized water and dried using nitrogen gas.
Diffuse reflection measurements were performed using an Agilent Cary 60 spectrometer, both
when the wafers were flat and after creating roughness. All traces were corrected for baseline
and scanned within the range of 300-800 nm. A white Polytetrafluoroethylene (PTFE) sample
was used to establish the 100$\%$ reflectance baseline. These measurements provided a
comparative analysis of the diffuse reflection spectra before and after the etching process,
highlighting the changes in surface properties due to the induced roughness.
Atomic Force Microscopy (AFM) measurements were conducted (see Fig. S.3) using an LS-AFM atomic force
microscope (AFM Workshop, USA) operating in tapping mode in the air at room temperature.
Silicon cantilever tips ACLA-10-W with reflective Al (back side) coating (Budget Sensors,
Innovative Solutions Ltd, Bulgaria) was used. Subsequently, all the images were flattened by
means of the AFM Control software. The software used for topographic data results
analysis and particle size determination is Gwyddion 2.60. The AFM analysis confirmed that the
surface roughness of the etched silicon wafers was consistently within the desired range of 8-9
nm.



\begin{figure}
\begin{tabular}{c}
    \begin{overpic}[width=0.9\columnwidth,keepaspectratio]{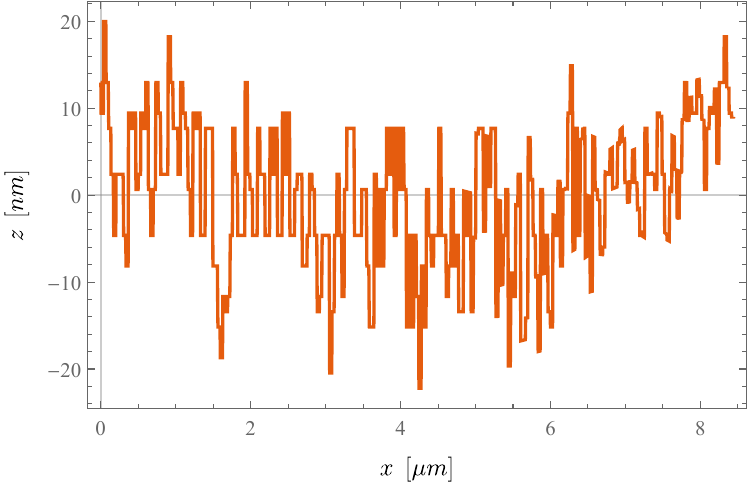}
    \put(0,0.56\columnwidth){(a)}
\end{overpic} \\
    \begin{overpic}[width=0.9\columnwidth,keepaspectratio]{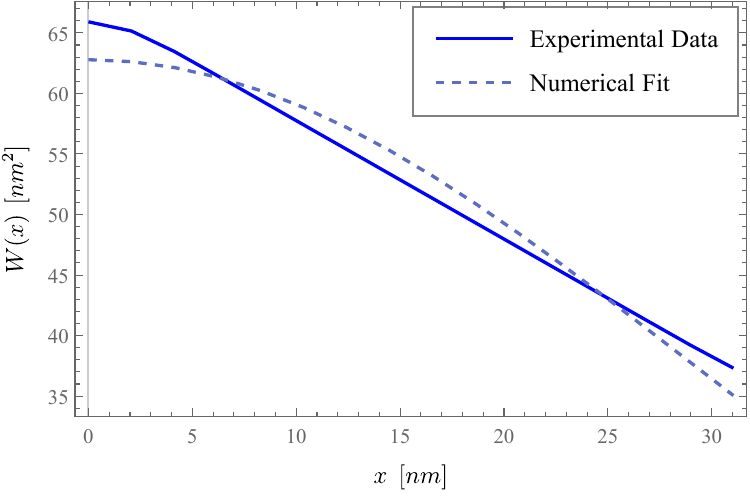}
    \put(0,0.56\columnwidth){(b)}
\end{overpic}\\
\end{tabular}
\caption{The experimentally obtained profile of the rough surface (a) and the calculated correlation function with fitting by the Gaussian function (b).}
\label{RoughProfAndCorr}
\end{figure}



In Fig. \ref{RoughProfAndCorr}, we illustrate our analysis with two plots. In Fig. \ref{RoughProfAndCorr} (a), we depict the roughness profile along a specific line, followed by the computation of its correlation function using the method detailed in \cite{bennett1976}, according to Eq. (S.62). \color{black} Fig. \ref{RoughProfAndCorr} (b) showcases the fitted results of this correlation function using the parameters \(\delta\) and \(a\) from the function \(W(x) = \delta^2 e^{-x^2/2a^2}\).

Subsequently, with the obtained fit parameters:

\begin{eqnarray}
    \delta=\text{(}7.92383 \pm 0.0438912\text{) nm} \nonumber\\
    a=\text{(}28.7482 \pm 0.751428\text{) nm}
    \label{fitparams}
\end{eqnarray}

we proceed to simulate the resulting reflection coefficient according to various theories. 

\begin{figure}
    \centering
    \includegraphics[width=0.9\columnwidth]{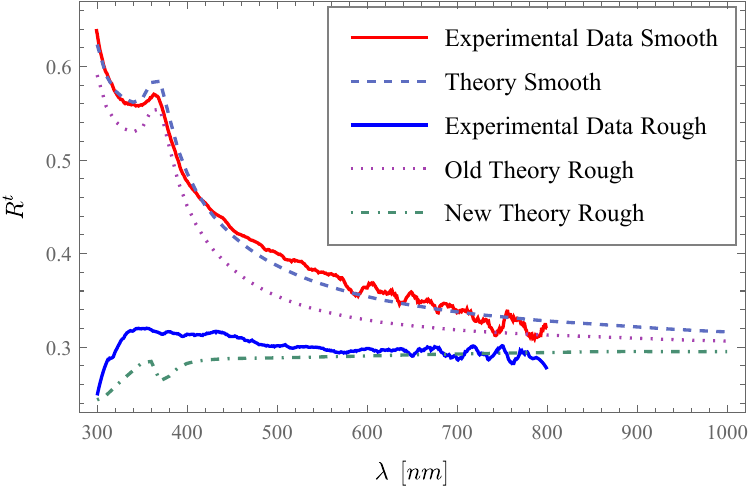}
    \caption{Comparison of the old perturbation theory by Navarrete \textit{et al.} (2009) \cite{navarrete2009} and the new one with experimental data.}
    \label{OldVsNew}
\end{figure}

To compare the theory presented in this paper with that in Navarrete \textit{et al.} (2009) \cite{navarrete2009}, we conduct a comparative analysis. We plot the results in Fig. \ref{OldVsNew}. 

First, we verify that the data from \cite{green2008} accurately reproduces the scattering results from the rough surface, ensuring consistency with experimental findings (see "Experimental Data Smooth" and "Theory Smooth"). Subsequently, we simulate the reflection coefficient for both theories for a rough surface using parameters as described in Eq. (\ref{fitparams}) (see "Old Theory Rough" and "New Theory Rough"). According to Fig. \ref{OldVsNew}, the predictions of the new theory exhibit a closer alignment with experimental observations (see "Experimental Data Rough").

\begin{figure}
\begin{tabular}{c}
    \begin{overpic}[width=0.9\columnwidth,keepaspectratio]{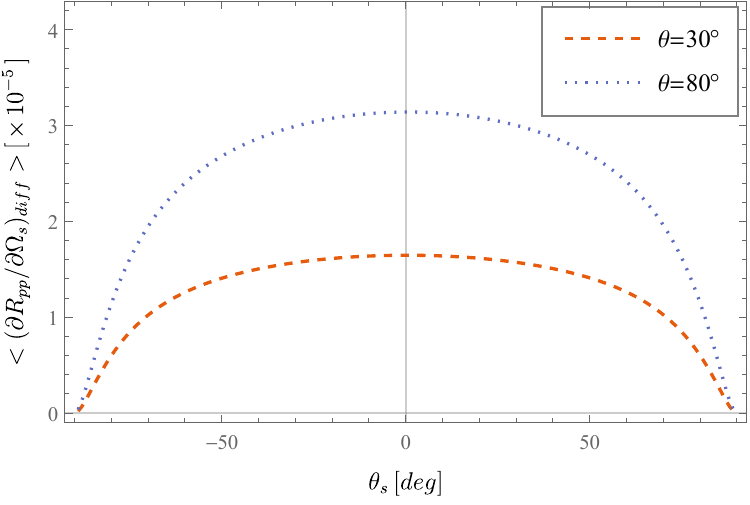}
    \put(0,0.56\columnwidth){(a)}
\end{overpic} \\
    \begin{overpic}[width=0.9\columnwidth,keepaspectratio]{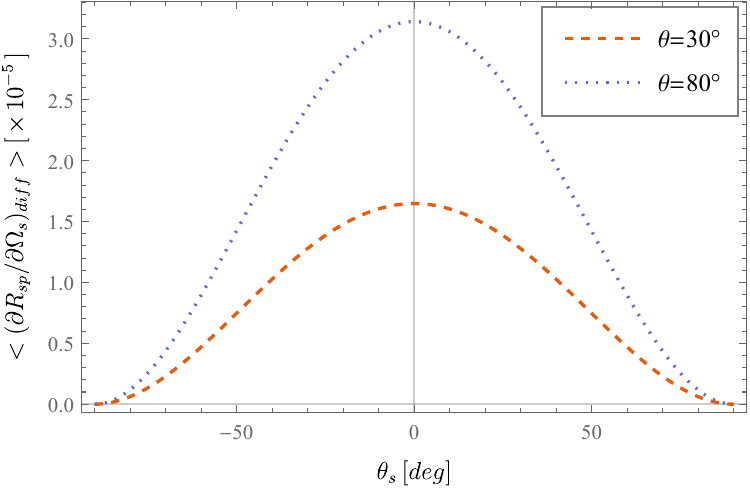}
    \put(0,0.56\columnwidth){(b)}
\end{overpic}\\
    \begin{overpic}[width=0.9\columnwidth,keepaspectratio]{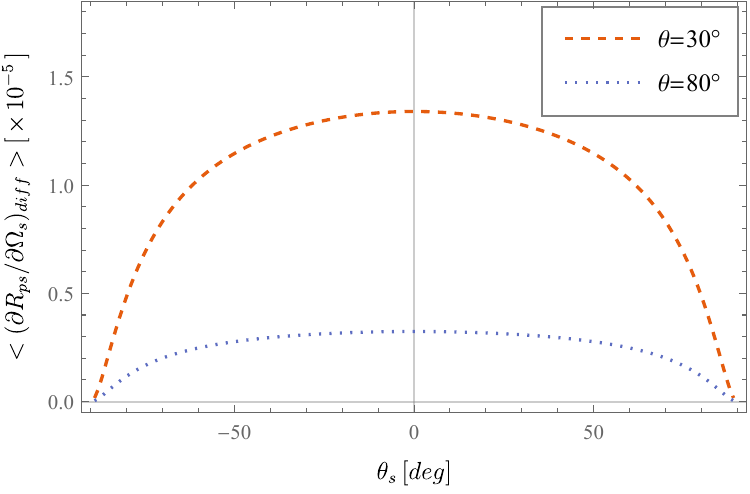}
    \put(0,0.56\columnwidth){(c)}
\end{overpic} \\
    \begin{overpic}[width=0.9\columnwidth,keepaspectratio]{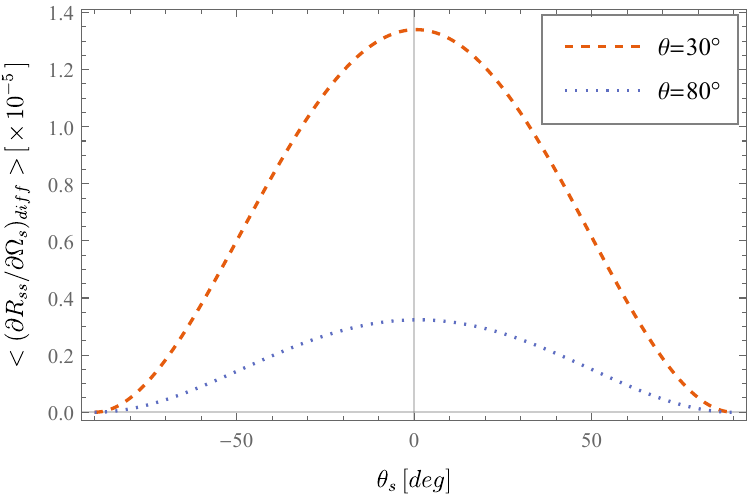}
    \put(0,0.56\columnwidth){(d)}
\end{overpic}\\
\end{tabular}
\caption{The numerically obtained sectional plots show the differential reflection coefficients for the diffuse components of beams reflected as P-polarized (a, c) and S-polarized (b, d) waves. These plots correspond to P-polarized (a, b) and S-polarized (c, d) incidences at a wavelength of 350 nm, varying with the scattering angle $\theta_s$. The incidences are depicted at 30$\degree$ (dashed line) and 80$\degree$ (dotted line). The correlation length is set to 10 nm, and the RMS roughness is 2 nm. In graphs (a, d), the cross-section is along the xOz plane, while in graphs (b, c), it is along the yOz plane.}
\label{DiffuseIntR}
\end{figure}

\textit{Results}-Fig. \ref{RoughVsFlat} showcases some potential anti-reflection effects induced solely by surface roughness. The total reflection coefficient experiences at least a fourfold reduction for incident p-polarized waves. We utilize dielectric constant data from \cite{green2008}. Figs. \ref{RoughVsFlat}(a, b) demonstrate the presence of anti-reflection across a broad range of incidence angles for each polarization. Fig. \ref{RoughVsFlat}(c) illustrates that the most significant anti-reflection occurs within the wavelength range of 300-400 nm.


\textit{Specular}-This effect of anti-reflection emerges primarily due to the reduction of the specular component of the reflection, while the diffuse component remains highly suppressed.

\textit{Haze}-Figs. \ref{DiffuseIntR} (a-d) illustrate the primary characteristics of diffuse scattering. The diffusely scattered light is predominantly p-polarized, a characteristic that remains independent of the incident polarization, and is directed along the surface normal, regardless of the angle of incidence. This represents the result of the interference from the multiple Rayleigh scatterings from each nanoscopic roughness on the surface in the limit $\delta, a \ll \lambda$. While each Rayleigh scattering from a single roughness should be isotropical, for their superposition only a particular direction becomes selected: the one along the normal to the surface. Maximum haze occurs for incident p-polarization at large angles, approaching Brewster's angle.

We can provide an explanation with use of another approximation. For $\beta \ll 1$, (according to the Supplemental Material) the haze intensity is approximately $k_0^4 \delta^2 a^2 \ll 1$, which is significantly smaller compared to the asymptotics given in Eq. (\ref{ApproxBigBeta}). This provides an estimate of the diffuse component intensities:

\begin{eqnarray}
\left<R^{diff}_{\alpha \beta}\right> \sim k_0^4 \delta^2 a^2 \sim 10^{-5},
\label{estimate}
\end{eqnarray}
where we use a roughness RMS $\delta = 2$ nm, correlation length $a = 10$ nm, and wavelength $\lambda = 350$ nm. These values align with those shown in Figs. \ref{DiffuseIntR} (a-d).

In Figs. \ref{DiffuseIntR} (a, b), which correspond to P-polarized incidence, we note the angular behavior of the scattered light. First, the maximal intensity corresponds to the direction along the normal to the surface and is independent of the angle of incidence. Second, the P-polarized scattering is spread wider over angles compared to the S-polarized one, which leads to higher integral intensity for P-polarized scattering. Third, the maximum of scattered intensity is achieved for the incidence angle, that is near to the Brewster's angle.

In Figs. \ref{DiffuseIntR} (c, d), which correspond to S-polarized incidence, we note, that only the third property is different. With the increase of the incidence angle, the reflection intensity declines.

Note, that this approximation for the specular part (in case of a normal incidence) results in:

\begin{eqnarray}
    \left<R^{spec}\right>=\left| \frac{\sqrt{\varepsilon}-1}{\sqrt{\varepsilon}+1} \right|^2\cdot \left[1 - 4\sqrt{2\pi}\frac{n}{|\varepsilon+1|^2}\frac{\delta^2}{ad}-8k_0^2\delta^2 n\right].\nonumber\\
\end{eqnarray}

An additional parameter comes to play, which was absent for transparent media: the skin depth $d$, which is given by

\begin{eqnarray}
    d=\frac{1}{\kappa k_0},
\end{eqnarray}

where $\kappa=Im(\sqrt{\varepsilon})$.

\color{black}

\color{black}

\textit{Summary}-We have examined specular and diffuse scattering from a rough, opaque surface. Our findings demonstrate that reducing specular reflection is achievable without significant haze generation, and we have established good agreement between theory and experiment.

\textit{Acknowledgement}-The authors express their gratitude to Armen Allahverdyan (Alikhanyan National Science Laboratory) for the conducive discussions.

This work was supported by the  Higher Education and Science Committee of MESCS RA,  in the frames of the research projects No. 21AG-1C062 and No. 23-2DP-1C010.



\begin{thebibliography}{99}
\bibitem{rad2020} A. S. Rad, A. Afshar, and M. Azadeh, Optical Materials {\bf 107}, 110027 (2020).
\bibitem{hossain2021} M. I. Hossain, B. A\"issa, A. Samara, S. A. Mansour, C. A. Broussillou, and V. B. Benito, ACS Omega {\bf 6}, 5276 (2021). 
\bibitem{isakov2020} K. Isakov, C. Kauppinen, S. Franssila, and H. Lipsanen, ACS Applied Materials $\&$amp; Interfaces {\bf 12}, 49957 (2020). 
\bibitem{wang2024} L. Wang, K. Liu, M. Yin, B. Yin, X. Liu, and S. Tang, Journal of Sol-Gel Science and Technology {\bf 109}, 835 (2024). 
\color{black}
\bibitem{chen2001}D. Chen, Solar Energy Materials and Solar Cells {\bf 68}, 313 (2001). 
\bibitem{ji2022}C. Ji, W. Liu, Y. Bao, X. Chen, G. Yang, B. Wei, F. Yang, and X. Wang, Photonics {\bf 9}, 906 (2022). 
\bibitem{chen2000}D. Chen, Y. Yan, E. Westenberg, D. Niebauer, N. Sakaitani, S. R. Chaudhuri, Y. Sato, and M. Takamatsu, Journal of Sol-Gel Science and Technology {\bf 19}, 77 (2000). 
\bibitem{mehmood2016}U. Mehmood, F. A. Al-Sulaiman, B. S. Yilbas, B. Salhi, S. H. A. Ahmed, and M. K. Hossain, Solar Energy Materials and Solar Cells {\bf 157}, 604 (2016). 
\bibitem{chen2011}J. Y. Chen and K. W. Sun, Thin Solid Films {\bf 519}, 5194 (2011). 
\bibitem{kim2013}K.-H. Kim and Q-Han Park, Scientific Reports {\bf 3}, (2013). 


\bibitem{amra1986}C. Amra, G. Albrand, and P. Roche, Applied Optics {\bf 25}, 2695 (1986). 
\bibitem{poitras2004}D. Poitras and J. A. Dobrowolski, Applied Optics {\bf 43}, 1286 (2004). 
\bibitem{moys1974}B. A. Moys, Thin Solid Films {\bf 21}, 145 (1974). 

\bibitem{kozhevnikov2024} I. V. Kozhevnikov, Q. Huang, Y. Zhuang, Z. Zhang, and Z. Wang, Optics Communications 130318 (2024). 
\bibitem{ong2024} Z.-Y. Ong, Physical Review B {\bf 109}, (2024). 
\color{black}

\bibitem{simonsen2010}I. Simonsen, The European Physical Journal Special Topics {\bf 181}, 1 (2010). 
\bibitem{hormann2023} S. M. H\"ormann, J. W. Hinum-Wagner, and A. Bergmann, Journal of Lightwave Technology {\bf 41}, 1503 (2023).
\bibitem{lamare2023} M. L. Lamare, J. D. Hedley, and M. D. King, The Cryosphere {\bf 17}, 737 (2023). 
\bibitem{ciochon2023} A. Ciochon, J. Kennedy, R. Leiba, L. Flanagan, and M. Culleton, Journal of Sound and Vibration {\bf 546}, 117434 (2023). 
\bibitem{hanson2024} S. Hanson, M. Takeda, W. Wang, and N. Ma, Effects of Surface Roughness of a Wave Plate on the Development of Diffractive Polarization Speckle (2024). 
\bibitem{ballington2024} H. Ballington and E. Hesse, Journal of Quantitative Spectroscopy and Radiative Transfer {\bf 323}, 109054 (2024). 
\color{black}
\bibitem{gevorkian2022}Z. S. Gevorkian, L. S. Petrosyan, and T. V. Shahbazyan, Physical Review B {\bf 106}, (2022). 
\bibitem{simonsen2009}I. Simonsen, \AA{}. Larsen, E. Andreassen, E. Ommundsen, and K. Nord-Varhaug, Physical Review A {\bf 79}, (2009). 
\bibitem{astm2000}ASTM Standard D 1003-95: Standard Test Method for Haze
and Luminous Transmittance of Transparent Plastics, 2000.
\bibitem{billmeyer1985}F. W. Billmeyer and Y. Chen, Color Res. Appl. {\bf 10}, 219 (1985).
\bibitem{mori2020}K. Mori, S. Samata, N. Mitsugi, A. Teramoto, R. Kuroda, T. Suwa, K. Hashimoto, and S. Sugawa, Japanese Journal of Applied Physics {\bf 59}, (2020). 
\bibitem{wang2005}J. Wang, E. Polizzi, A. Ghosh, S. Datta, and M. Lundstrom, Applied Physics Letters {\bf 87}, (2005). 
\bibitem{khan2007}H. Khan, D. Mamaluy, and D. Vasileska, Journal of Vacuum Science $\&$amp; Technology B: Microelectronics and Nanometer Structures Processing, Measurement, and Phenomena {\bf 25}, 1437 (2007). 
\bibitem{carminati2020}J.-P. Banon, I. Simonsen, and R. Carminati, Physical Review A {\bf 101}, (2020).  

\bibitem{moreno2010}M. Moreno, D. Daineka, and P. Roca i Cabarrocas, Solar Energy Materials and Solar Cells {\bf 94}, 733 (2010). 
\bibitem{armstrong2018}A. Godoy, F. G. Carlucci, D. M. Leite, W. Miyakawa, A. L. Pereira, M. Massi, and A. S. da Silva Sobrinho, Surface and Coatings Technology {\bf 354}, 153 (2018).  
\bibitem{stepf2017}A. Stapf, F. Honeit, C. Gondek, and E. Kroke, Solar Energy Materials and Solar Cells {\bf 159}, 112 (2017).  
\bibitem{zrir2024}M. A. Zrir, M. Kakhia, and N. AlKafri, Journal of Materials Science {\bf 59}, 2048 (2024).  
\bibitem{dimitrov2013}D. Z. Dimitrov and C.-H. Du, Applied Surface Science {\bf 266}, 1 (2013).  
\bibitem{pasanen2020}T. P. Pasanen, J. Isomets\"a, M. Garin, K. Chen, V. V\"ah\"anissi, and H. Savin, Advanced Optical Materials {\bf 8}, (2020).  
\bibitem{islam2015}M. Islam, A. Sajid, M. A. Mahmood, M. M. Bellah, P. B. Allen, Y.-T. Kim, and S. M. Iqbal, Nanotechnology {\bf 26}, 225101 (2015).  

\bibitem{gevorkian2024} V. Gareyan and Zh. Gevorkian, Physical Review A {\bf 109}, (2024). 
\bibitem{navarrete2009}A. G. Navarrete Alcal\'a, E. I. Chaikina, E. R. M\'endez, T. A. Leskova, and A. A. Maradudin, Waves in Random and Complex Media {\bf 19}, 600 (2009).
\bibitem{bennett1976}J. M. Bennett, Applied Optics {\bf 15}, 2705 (1976). 
\bibitem{green2008}M. A. Green, Solar Energy Materials and Solar Cells {\bf 92}, 1305 (2008). 
\end{thebibliography}
\end{document}


\title{- Supplemental Material -\\ Nanoroughness induced anti-reflection and haze effects in opaque systems}
\author{V.Gareyan$^{1}$, N.Margaryan$^{1}$ and Zh.Gevorkian$^{1,2}$}
\affiliation{$^{1}$ Alikhanyan National Laboratory, Alikhanian Brothers St. 2,  Yerevan 0036 Armenia\\
$^{2}$ Institute of Radiophysics and Electronics, Ashtarak-2 0203 Armenia}

\maketitle

\tableofcontents

\renewcommand{\theequation}{S.\arabic{equation}}
\setcounter{equation}{0}
\renewcommand{\thefigure}{S.\arabic{figure}}
\setcounter{figure}{0}

\section{Note 1: Theory}

\subsection{1.1 Reflection Coefficients}

\subsubsection{1.1.1 Transversality of waves}


We start from the wave equation:

\begin{eqnarray}
    \nabla^2 \textbf{E} - \nabla (\nabla \cdot \textbf{E}) + k_0^2 \varepsilon(\boldsymbol{\rho},z) \textbf{E} = 0,
\end{eqnarray}

where the effective dielectric constant  $\varepsilon(\boldsymbol{\rho},z)=\varepsilon_0(\boldsymbol{\rho},z)+\varepsilon_1(\boldsymbol{\rho},z)$ and 

\begin{align}
\varepsilon_0(\boldsymbol{\rho},z)=\left\{\begin{array}{cc}
         1, \mbox{~~~~for~} z>h(\boldsymbol{\rho}) \\
         \varepsilon, \mbox{~~~~for~} z<h(\boldsymbol{\rho})
         \end{array}\right.
\label{epszero}
\end{align}

together with

\begin{equation}
\varepsilon_1(\boldsymbol{\rho},z)=[\varepsilon-1]\times \delta(z-h(\boldsymbol{\rho})) \times h(\boldsymbol{\rho}).
\label{eps1}
\end{equation}

are already shifted to the actual boundary \cite{carminati,gevorkian22}, where we denote $\varepsilon(\omega) \equiv \varepsilon$ for brevity (see  in \cite{gevorkian2024}).

First, consider the case of the incident wave with p-polarization. We will show that only in this case the reflection coefficients become quite sensitive to the change of the system parameters, which opens a perspective for these effects in terms of a practical usage. The strength of the electric component of the reference field is 
$ \textbf{E}_{0p} = \textbf{e}_x E_{0p,x} + \textbf{e}_z E_{0p,z}$, where

\begin{align}
E_{0p,x}({\bf r})=\left\{\begin{array}{c}\cos\theta \times (e^{ik_xx-ik_{z}z}-r_p e^{ik_xx+ik_{z}z}),\\ \mbox{~for~} z>h(x,y),
\\ \\
t_p\cos\theta_t \times e^{ik_xx+ik_{-}z}, \\ \mbox{\qquad\qquad for~} z<h(x,y),
\end{array}\right.
\label{backx}
\end{align}
and
\begin{align}
E_{0p,z}({\bf r})=\left\{\begin{array}{c}\sin\theta \times(e^{ik_xx-ik_{z}z}+r_p e^{ik_xx+ik_{z}z}),\\ \mbox{~for~} z>h(x,y),
\\ \\
t_p \sin\theta_t \times e^{ik_xx+ik_{-}z}, \\ \mbox{\qquad\qquad for~} z<h(x,y),
\end{array}\right.
\label{backz}
\end{align}
and $\sin\theta_t=\sin\theta/\sqrt{\varepsilon}$, $ \cos\theta_t=\sqrt{1-\sin^2\theta/\varepsilon}$, $\theta_t$ is the transmission angle and is generally complex and $\break k_{-}=-k_0 \sqrt{\varepsilon-\sin^2\theta}$, $k_x=k_0 \sin\theta$, $k_z=k_0 \cos\theta$ and $r_p$ and $t_p$ are the reflection and transmission coefficients from a smooth surface for a p-polarized wave:

\begin{eqnarray}
 r_p=\frac{\varepsilon \cos\theta- \sqrt{\varepsilon-\sin^2\theta}}{\varepsilon \cos\theta+\sqrt{\varepsilon-\sin^2\theta}}, \nonumber\\
 t_p=\frac{2\sqrt{\varepsilon}\cos\theta}{\varepsilon \cos\theta+\sqrt{\varepsilon-\sin^2\theta}}.
\label{ppolar}
\end{eqnarray}

The perturbation theory gives the solution of the problem in a form $\textbf{E} = \sum_{n=0}^{\infty} \textbf{E}_s^{(n)}$, where $\textbf{E}_s^{(0)}=\textbf{E}_0$ is the reference field and

\begin{eqnarray}
    \textbf{E}_s^{(n+1)}({\bf r}) = -\frac{k_0^2}{4\pi} \int d\textbf{r}'\textbf{D}_0({\bf r}, {\bf r}')\textbf{E}_s^{(n)}({\bf r}')\varepsilon_1({\bf r}')
\end{eqnarray}

for any $n=0,1,2,...$, where

\begin{eqnarray}
    \textbf{D}_0({\bf r}, {\bf r}') = \int \frac{d\textbf{q}}{(2\pi)^2} \textbf{d} (\textbf{q},z,z')e^{i\textbf{q}(\boldsymbol{\rho}-\boldsymbol{\rho}')}
\end{eqnarray}

is the dyadic Green's function for the case of the flat interface with 

\begin{gather}
    \textbf{d} (\textbf{q},z,z') =\nonumber\\ = \frac{1}{q^2}\left[
    \begin{array}{ccc}
        q_x^2 g_{xx} + q_y^2 g_{yy} & q_x q_y (g_{xx} - g_{yy}) & q q_x g_{xz} \\
        q_x q_y (g_{xx} - g_{yy}) & q_y^2 g_{xx} + q_x^2 g_{yy} & q q_y g_{xz} \\
        q q_x g_{zx} & q q_y g_{zx} & q^2 g_{zz}
    \end{array}
    \right]_{(\textbf{q},z,z')}.
    \label{dqzz}
\end{gather}

The total electric field in the vacuum ($z>h(\boldsymbol{\rho})$) can be decomposed into the incident and scattered parts: $\textbf{E} = \textbf{E}_{inc} + \textbf{E}_{sc}$. For p-polarized incident wave we have 

\begin{eqnarray}
    \textbf{E}_{inc}(\boldsymbol{\rho},z) = (\textbf{e}_x \cos\theta + \textbf{e}_z \sin\theta) \times e^{i k_x x - i k_z z}
\end{eqnarray}

and

\begin{gather}
    \textbf{E}_{sc}(\boldsymbol{\rho},z) = r_p\times (-\textbf{e}_x \cos\theta + \textbf{e}_z \sin\theta) \times e^{i k_x x + i k_z z} +\nonumber\\+ \left(-\frac{k_0^2}{4\pi}\right) \int d\textbf{r}'\textbf{D}_0({\bf r}, {\bf r}')\textbf{E}_p({\bf r}')\varepsilon_1({\bf r}'),
    \label{scatp}
\end{gather}

where we used ${\bf r} = \boldsymbol{\rho} + \textbf{e}_z z$. The second term on the r.h.s. of Eq.(\ref{scatp}) also describes a transversal wave. To prove this we note that 1) in practice the observation point is always above the rough surface i.e. it is enough to consider the case $z > max[h(\boldsymbol{\rho})]$ and 2) because of $\varepsilon_1$ the integration over $z'$ reduces to the rough interface. Thus, we can consider the case $z>z'$, where we have the relations 

\begin{eqnarray}
    g_{z\beta}= -\frac{q}{k(q)} g_{x\beta} \nonumber\\
    \frac{\partial}{\partial z} g_{\alpha \beta}= i k(q) g_{\alpha \beta}
\end{eqnarray}

for $\alpha,\beta = x$ or $z$, derived from their definitions in \cite{maradudin1976}. Now, to prove the transversality of $\textbf{E}_{sc}$, we need to consider its divergence:

\begin{eqnarray}
    \nabla \cdot \textbf{E}_{sc}=-i\frac{k_0^2}{4\pi}\int d{\bf r}'\int \frac{d\boldsymbol{q}}{(2\pi)^2} e^{i\boldsymbol{q}(\boldsymbol{\rho}-\boldsymbol{\rho}')}\times \nonumber\\ \times(\boldsymbol{q} + k(q)\textbf{e}_z)^T \textbf{d}({\bf q},z,z')\textbf{E}_p({\bf r'})\varepsilon_1({\bf r'})
\end{eqnarray}

and, specifically, the dot product

\begin{gather}
    (\boldsymbol{q} + k(q)\textbf{e}_z)^T \textbf{d}({\bf q},z,z') = \frac{1}{q^2} (q_x,q_y,k(q))\cdot \nonumber\\ \cdot\left(
    \begin{array}{ccc}
        q_x^2 g_{xx} + q_y^2 g_{yy} & q_x q_y (g_{xx} - g_{yy}) & q q_x g_{xz} \\
        q_x q_y (g_{xx} - g_{yy}) & q_y^2 g_{xx} + q_x^2 g_{yy} & q q_y g_{xz} \\
        q q_x g_{zx} & q q_y g_{zx} & q^2 g_{zz}
    \end{array}
    \right)_{(\textbf{q},z,z')} \nonumber\\ = (0, 0, 0)
\end{gather}

which proves that $\nabla\cdot\textbf{E}_{sc}=0$ or the transversality of the scattered wave. 

\subsubsection{1.1.2 Further derivation}

This makes available writing the scattered wave in a form

\begin{gather}
    \textbf{E}_{sc}(\boldsymbol{\rho},z)= \int \frac{d\boldsymbol{q}}{(2\pi)^2}e^{i\boldsymbol{q}\boldsymbol{\rho} + i k(q) z}\nonumber\\ \left[ \frac{(\hat{\boldsymbol{q}}k(q)-\textbf{e}_z q)}{k_0} A_p(\boldsymbol{q}) + (\textbf{e}_z \times \hat{\boldsymbol{q}}) A_s(\boldsymbol{q}) \right] 
\end{gather}

with scattered p- and s-polarized wave amplitudes

\begin{gather}
    A_p(\boldsymbol{q})=-r_p(2\pi)^2\delta(\boldsymbol{q}-k_x\textbf{e}_x)-\frac{k_0^2}{4\pi}\int d{\bf r}' e^{-i\boldsymbol{q}\boldsymbol{\rho}'}\times \nonumber\\ \times\frac{k_0}{k(q)}(\hat{q}_x g_{xx}^{0+}, \hat{q}_y g_{xx}^{0+}, g_{xz}^{0+})\cdot \textbf{E}_p({\bf r'})\varepsilon_1({\bf r'}),\nonumber\\
    A_s(\boldsymbol{q})=-\frac{k_0^2}{4\pi}\int d{\bf r}' e^{-i\boldsymbol{q}\boldsymbol{\rho}'}\times\nonumber\\ \times\frac{k_0}{k(q)}(-\hat{q}_y g_{yy}^{0+}, \hat{q}_x g_{yy}^{0+}, 0)\cdot \textbf{E}_p({\bf r'})\varepsilon_1({\bf r'})
\end{gather}

, where we introduced the notation 

\begin{eqnarray}
    g_{\alpha \beta}^{0+}(\boldsymbol{q},z') = e^{-i k(q)z} g_{\alpha \beta}(\boldsymbol{q},z,z').
\end{eqnarray}

According to \cite{navarrete2009} we can write 

\begin{eqnarray}
    A_{\alpha}(\boldsymbol{q})=\sum_{\beta=p,s} R_{\alpha \beta}(\boldsymbol{q}|\textbf{e}_x k_x) B_{\beta}(\textbf{e}_x k_x) =\nonumber\\= -R_{\alpha p}(\boldsymbol{q}|\textbf{e}_x k_x),
\end{eqnarray}
since $B_s=0$ and $B_p=-1$ for p-polarized incidence. Thus, finally, we can write the reflection coefficients as

\begin{gather}
    R_{pp}(\boldsymbol{q}|\textbf{e}_x k_x)=r_p(2\pi)^2\delta(\boldsymbol{q}-k_x\textbf{e}_x)+\frac{k_0^2}{4\pi}\int d{\bf r}' e^{-i\boldsymbol{q}\boldsymbol{\rho}'}\times \nonumber\\ \times\frac{k_0}{k(q)}(\hat{q}_x g_{xx}^{0+}, \hat{q}_y g_{xx}^{0+}, g_{xz}^{0+})\cdot \textbf{E}_p({\bf r'})\varepsilon_1({\bf r'}),\nonumber\\
    R_{sp}(\boldsymbol{q}|\textbf{e}_x k_x)=\frac{k_0^2}{4\pi}\int d{\bf r}' e^{-i\boldsymbol{q}\boldsymbol{\rho}'}\times \nonumber\\ \times\frac{k_0}{k(q)}(-\hat{q}_y g_{yy}^{0+}, \hat{q}_x g_{yy}^{0+}, 0)\cdot \textbf{E}_p({\bf r'})\varepsilon_1({\bf r'}).
    \label{RppRsp}
\end{gather}

The differential reflection coefficient

\begin{eqnarray}
    \frac{\partial R_{\alpha \beta}}{\partial \Omega_s} = \frac{1}{S_0}\frac{1}{\lambda^2}\frac{\cos^2\theta_s}{\cos\theta}  |R_{\alpha \beta}(\boldsymbol{q}|\textbf{e}_x k_x)|^2,
\end{eqnarray}

where $\boldsymbol{q}=k_0\sin\theta_s(\textbf{e}_x\cos\phi_s + \textbf{e}_y\sin\phi_s)$, now, can be represented as a sum of specular and diffuse reflection parts. Indeed, the square of the absolute value of each reflection coefficient in Eq.(\ref{RppRsp})

\begin{gather}
    |R_{pp}(\boldsymbol{q}|\textbf{e}_x k_x)|^2=\left[ |r_p|^2  +  \frac{k_0^2}{2\pi S_0}\times\right.\nonumber\\\left. \times Re\left( 
\frac{r_p^*}{\cos \theta} \int d{\bf r}' e^{-ik_x x'} (g_{xx}^{0+},0, g_{xz}^{0+})|_{(\boldsymbol{e}_x k_x, z')}\cdot \textbf{E}_p({\bf r'})\varepsilon_1({\bf r'}) \right)\right]\times \nonumber\\ \times S_0(2\pi)^2\delta(\boldsymbol{q}-\textbf{e}_x k_x) + \left(\frac{k_0^2}{4\pi}\right)^2 \left| \int d{\bf r}' e^{-i\boldsymbol{q}\boldsymbol{\rho}'}\times\right.\nonumber\\ \left.\times\frac{k_0}{k(q)}(\hat{q}_x g_{xx}^{0+}, \hat{q}_y g_{xx}^{0+}, g_{xz}^{0+})\cdot \textbf{E}_p({\bf r'})\varepsilon_1({\bf r'})\right|^2,\nonumber\\
|R_{sp}(\boldsymbol{q}|\textbf{e}_x k_x)|^2=\left(\frac{k_0^2}{4\pi}\right)^2\left|\int d{\bf r}' e^{-i\boldsymbol{q}\boldsymbol{\rho}'}\times\right.\nonumber\\ \left.\times\frac{k_0}{k(q)}(-\hat{q}_y g_{yy}^{0+}, \hat{q}_x g_{yy}^{0+}, 0)\cdot \textbf{E}_p({\bf r'})\varepsilon_1({\bf r'})\right|^2
\label{Rsquareppsp}
\end{gather}

consists of two summands: with $\delta$-function (which is the specular part) and without it (which is the diffuse part). This can be written as $|R_{\alpha \beta}(\boldsymbol{q}|\textbf{e}_x k_x)|^2=|R_{\alpha \beta}(\boldsymbol{q}|\textbf{e}_x k_x)|^2_{spec}+|R_{\alpha \beta}(\boldsymbol{q}|\textbf{e}_x k_x)|^2_{diff}$. It is important to derive the expressions for the integral reflection coefficients:

\begin{gather}
    {\bf R}_{\alpha \beta}=\int d\Omega_s \frac{\partial R_{\alpha \beta}}{\partial \Omega_s}=\nonumber\\=\frac{1}{S_0 \cos \theta} \int_{q<k_0} \frac{d\boldsymbol{q}}{(2\pi)^2} \frac{k(q)}{k_0} |R_{\alpha \beta}(\boldsymbol{q}|\textbf{e}_x k_x)|^2
\end{gather}

This quantity can also be represented as a sum of the scattering and the diffuse parts, which can be written as follows: ${\bf R}_{\alpha \beta}={\bf R}_{\alpha}^{spec} \delta_{\alpha \beta} + {\bf R}_{\alpha \beta}^{diff}$, where we used the fact (which we will prove later also in the case of incident s-polarization or $\beta = s$), that the off-diagonal elements in the specular part of the reflection matrix $||{\bf R}_{\alpha \beta}||$ vanish. In our case

\begin{gather}
    {\bf R}_{p}^{spec}=|r_p|^2  +  \frac{k_0^2}{2\pi S_0}Re\left(r_p^* 
\frac{1}{\cos \theta} \int d{\bf r}' e^{-ik_x x'}\times\right.\nonumber\\ \left.\times(g_{xx}^{0+},0, g_{xz}^{0+})|_{(\boldsymbol{e}_x k_x, z')}\cdot \textbf{E}_p({\bf r'})\varepsilon_1({\bf r'})\right),\nonumber\\
{\bf R}_{pp}^{diff}=\left(\frac{k_0^2}{4\pi}\right)^2 \frac{1}{S_0 \cos \theta} \int_{q<k_0} \frac{d\boldsymbol{q}}{(2\pi)^2} \frac{k(q)}{k_0} \left| \int d{\bf r}' e^{-i\boldsymbol{q}\boldsymbol{\rho}'}\frac{k_0}{k(q)}\times\right.\nonumber\\ \left.\times(\hat{q}_x g_{xx}^{0+}, \hat{q}_y g_{xx}^{0+}, g_{xz}^{0+})\cdot \textbf{E}_p({\bf r'})\varepsilon_1({\bf r'})\right|^2,\nonumber\\
{\bf R}_{sp}^{diff}={\bf R}_{sp}=\left(\frac{k_0^2}{4\pi}\right)^2 \frac{1}{S_0 \cos \theta} \int_{q<k_0} \frac{d\boldsymbol{q}}{(2\pi)^2} \frac{k(q)}{k_0} \times\nonumber\\ \times \left|\int d{\bf r}' e^{-i\boldsymbol{q}\boldsymbol{\rho}'}\frac{k_0}{k(q)}(-\hat{q}_y g_{yy}^{0+}, \hat{q}_x g_{yy}^{0+}, 0)\cdot \textbf{E}_p({\bf r'})\varepsilon_1({\bf r'})\right|^2.
\label{SpecAndDiffppsp}
\end{gather}

However, in practice, we need to consider quantities that are averaged over the ensemble of realization of the surface profile $h(\boldsymbol{\rho})$ with correlation function $\left<h(\boldsymbol{\rho})h(\boldsymbol{\rho}')\right>=\delta^2e^{-(\boldsymbol{\rho}-\boldsymbol{\rho}')^2/2a^2}$:

\begin{eqnarray}
    \left< \frac{\partial R_{\alpha \beta}}{\partial \Omega_s} \right> = \frac{1}{S_0}\frac{1}{\lambda^2}\frac{\cos^2\theta_s}{\cos\theta} \left< |R_{\alpha \beta}(\boldsymbol{q}|\textbf{e}_x k_x)|^2\right>.
\end{eqnarray}

Moreover, dividing the quantities into the coherent and incoherent parts can also be useful. For the differential reflection coefficient, this can be done as follows:

\begin{gather}
    \left< \frac{\partial R_{\alpha \beta}}{\partial \Omega_s} \right>_{coh} = \frac{1}{S_0}\frac{1}{\lambda^2}\frac{\cos^2\theta_s}{\cos\theta} |\left< R_{\alpha \beta}(\boldsymbol{q}|\textbf{e}_x k_x)\right>|^2 \nonumber\\
    \left< \frac{\partial R_{\alpha \beta}}{\partial \Omega_s} \right>_{incoh} = \frac{1}{S_0}\frac{1}{\lambda^2}\frac{\cos^2\theta_s}{\cos\theta}\times\nonumber\\ \times \left[\left< |R_{\alpha \beta}(\boldsymbol{q}|\textbf{e}_x k_x)|^2\right>-|\left< R_{\alpha \beta}(\boldsymbol{q}|\textbf{e}_x k_x)\right>|^2\right].
\end{gather}

We will show that, within the framework of second-order perturbation theory, the coherent and incoherent scattering parts correspond to the specular and diffuse scattering respectively. 


To proceed, we need to make assumptions regarding any given magnitude on the rough interface. Typically, this magnitude is assumed to be equal to the mean of the respective limiting values on both sides of the interface. However, if the magnitudes on both sides are of the same order, we can instead use the limiting value from either side. Mathematically, this can be expressed as:

\begin{equation}
\varepsilon_1(\boldsymbol{\rho},z)=[\varepsilon-1]\times \delta(z-h^{-}(\boldsymbol{\rho})) \times h(\boldsymbol{\rho}).
\label{eps1'}
\end{equation}

First consider the diffuse part.


Substituting Eq. (\ref{eps1'}) into Eq. (\ref{Rsquareppsp}), and then averaging, followed by the use of the identity

\begin{eqnarray}
    \int d\boldsymbol{\rho}' \int d\boldsymbol{\rho}'' \left< h(\boldsymbol{\rho}') h(\boldsymbol{\rho}'') \right> e^{i\boldsymbol{q}(\boldsymbol{\rho}'-\boldsymbol{\rho}'')} =\nonumber\\= 2\pi S_0 a^2 \delta^2 exp \left(-\frac{q^2 a^2}{2} \right)
\end{eqnarray}

we get

\begin{gather}
    \left<|R_{pp}(\boldsymbol{q}|\textbf{e}_x k_x)|^2_{diff}\right>=\nonumber\\=2\pi S_0 \frac{|\varepsilon - 1|^2}{\cos^2\theta_s}\delta^2
 |t_p|^2 e^{-\frac{1}{2}\beta^2(\sin^2 \theta + \sin^2 \theta_s)} e^{\beta^2\sin\theta\sin\theta_s\cos\phi_s} \times \nonumber\\ \times \left| G_{xx}^{(+-)}(r=\beta \sin \theta_s)\sqrt{1-\frac{\sin^2\theta}{\varepsilon}} \cos \phi_s +\right.\nonumber\\ \left. G_{xz}^{(+-)}(r=\beta \sin \theta_s) \frac{\sin \theta}{\sqrt{\varepsilon}} \right|^2 , \nonumber\\
\left<|R_{sp}(\boldsymbol{q}|\textbf{e}_x k_x)|^2_{diff}\right>=\left<|R_{sp}(\boldsymbol{q}|\textbf{e}_x k_x)|^2\right>=\nonumber\\= 2\pi S_0 |\varepsilon - 1|^2\delta^2
 |t_p|^2 e^{-\frac{1}{2}\beta^2(\sin^2 \theta + \sin^2 \theta_s)} e^{\beta^2\sin\theta\sin\theta_s\cos\phi_s} \times \nonumber\\ \times \left| G_{yy}^{(+-)}(r=\beta \sin \theta_s)\sqrt{1-\frac{\sin^2\theta}{\varepsilon}} \sin \phi_s \right|^2,
\end{gather}

where $\beta = k_0 a$ and

\begin{eqnarray}
    G_{xx}^{(+-)}(r)=\frac{\sqrt{\varepsilon\beta^2-r^2}\sqrt{\beta^2-r^2}}{\varepsilon \sqrt{\beta^2-r^2}+\sqrt{\varepsilon \beta^2-r^2}}\nonumber\\
    G_{xz}^{(+-)}(r)=-\frac{r \sqrt{\beta^2-r^2}}{\varepsilon \sqrt{\beta^2-r^2}+\sqrt{\varepsilon \beta^2-r^2}}\nonumber\\
    G_{yy}^{(+-)}(r)=\frac{\beta^2}{\sqrt{\beta^2-r^2}+\sqrt{\varepsilon \beta^2-r^2}}
\end{eqnarray}

Note that averaging over ensembles of realizations of nanorough surfaces, due to the smallness of \(\delta\), allows us to neglect terms of order \(O(\delta^4)\). Therefore, in this case, we only consider terms of order \(\delta^2\), which arise from the multiplication of two \(h\) terms inside each \(\varepsilon_1\) in the square of the absolute value of the integration function.

This leads to the expression for the differential diffuse reflection coefficient:

\begin{gather}
    \left<\left(\frac{\partial R_{pp}}{\partial \Omega_s}\right)_{diff}\right>=\nonumber\\=\frac{2\pi \delta^2}{\lambda^2} \frac{|\varepsilon - 1|^2}{\cos\theta}
 |t_p|^2 e^{-\frac{1}{2}\beta^2(\sin^2 \theta + \sin^2 \theta_s)}  e^{\beta^2\sin\theta\sin\theta_s\cos\phi_s} \times \nonumber\\ \times \left| G_{xx}^{(+-)}(r=\beta \sin \theta_s)\sqrt{1-\frac{\sin^2\theta}{\varepsilon}} \cos \phi_s +\right.\nonumber\\ \left. +G_{xz}^{(+-)}(r=\beta \sin \theta_s) \frac{\sin \theta}{\sqrt{\varepsilon}} \right|^2 , \nonumber\\
\left<\left(\frac{\partial R_{sp}}{\partial \Omega_s}\right)_{diff}\right>=\left<\left(\frac{\partial R_{sp}}{\partial \Omega_s}\right)\right>= \nonumber\\=
\frac{2\pi \delta^2}{\lambda^2}\frac{\cos^2\theta_s}{\cos\theta}  |\varepsilon - 1|^2
 |t_p|^2 e^{-\frac{1}{2}\beta^2(\sin^2 \theta + \sin^2 \theta_s)} e^{\beta^2\sin\theta\sin\theta_s\cos\phi_s} \times \nonumber\\ \times \left| G_{yy}^{(+-)}(r=\beta \sin \theta_s)\sqrt{1-\frac{\sin^2\theta}{\varepsilon}} \sin \phi_s \right|^2.
 \label{diffRdiffuseppsp}
\end{gather}

To obtain the integral diffuse reflection coefficient, we integrate Eq. (\ref{diffRdiffuseppsp}) over the entire upper hemisphere. We can simplify the integration over the azimuthal angle \(\phi_s\) using the definition of the modified Bessel function of order \(n\):

\begin{equation}
    I_n(x)=\int_{0}^{2\pi} \frac{d\phi_s}{2\pi} e^{x\cos\phi_s \pm i n\phi_s} = \int_{0}^{2\pi} \frac{d\phi_s}{2\pi} e^{x\cos\phi_s}\cos (n\phi_s).
\end{equation}

It is possible to change the integration variable and use $r=qa=\beta\sin\theta_s$ instead of $\theta_s$:

\begin{gather}
    \left<{\bf R}_{pp}^{diff}\right>=\frac{|\varepsilon-1|^2}{\cos\theta} \left(\frac{\delta}{a}\right)^2 |t_p|^2 e^{-\frac{1}{2}\beta^2\sin^2\theta} \times \nonumber\\ \times\int_0^{\beta}dr r \frac{\beta}{\sqrt{\beta^2-r^2}} e^{-\frac{1}{2} r^2}\times \nonumber\\ \times \left[ |G_{xx}^{(+-)}(r)|^2\cdot \left| 1-\frac{\sin^2\theta}{\varepsilon} \right|\cdot  \frac{I_0+I_2}{2}\Bigg|_{x=r\beta\sin\theta}+ \right.\nonumber\\ \left.+2 Re\left(\left[ G_{xx}^{(+-)}(r) \sqrt{ 1-\frac{\sin^2\theta}{\varepsilon} }  \text{ } \right]^{*}G_{xz}^{(+-)}(r)\frac{\sin\theta}{\sqrt{\varepsilon}}\right)\times \right.\nonumber\\ \left. \times I_1(r\beta\sin\theta)+ |G_{xz}^{(+-)}(r)|^2\frac{\sin^2\theta}{|\varepsilon|} I_0(r\beta\sin\theta) \right], \nonumber\\
    \left<{\bf R}_{sp}^{diff}\right>=\left<{\bf R}_{sp}\right>=\nonumber\\=\frac{|\varepsilon-1|^2}{\cos\theta} \left(\frac{\delta}{a}\right)^2 |t_p|^2 e^{-\frac{1}{2}\beta^2\sin^2\theta} \times \nonumber\\ \times \int_0^{\beta}dr r \frac{\sqrt{\beta^2-r^2}}{\beta} e^{-\frac{1}{2} r^2} |G_{yy}^{(+-)}(r)|^2\frac{I_0-I_2}{2}\Bigg|_{x=r\beta\sin\theta},
\end{gather}
where the argument for each $I_n$ ($n=0,1,2$) is $x=r\beta \sin\theta$. This form is more convenient for making the approximations described in the following sections.

Now, consider the specular part. Here, calculations become slightly more complicated because we need to expand Taylor series in powers of \( h \) for the function inside the integral. For the specular reflection coefficient in Eq. (\ref{SpecAndDiffppsp}), we can express it as:
\begin{widetext}
\begin{gather}
    \left<{\bf R}_p^{spec}\right>={\bf R}_{p,0}^{spec}+{\bf R}_{p,i1}^{spec}+{\bf R}_{p,i2}^{spec},\nonumber\\
    {\bf R}_{p,0}^{spec}=|r_p|^2, \nonumber\\
    {\bf R}_{p,i1}^{spec}=\frac{k_0^2}{2\pi S_0}Re\left(r_p^* 
\frac{1}{\cos \theta} \left< \int d{\bf r}' e^{-ik_x x'}(g_{xx}^{0+},0, g_{xz}^{0+})|_{(\boldsymbol{e}_x k_x, z')}\cdot \textbf{E}_{0p}({\bf r'})\varepsilon_1({\bf r'}) \right> \right) ,\nonumber\\
{\bf R}_{p,i2}^{spec}=\frac{k_0^2}{2\pi S_0}Re\left(r_p^* 
\frac{1}{\cos \theta} \left< \int d{\bf r}' e^{-ik_x x'}(g_{xx}^{0+},0, g_{xz}^{0+})|_{(\boldsymbol{e}_x k_x, z')}\cdot \textbf{E}_{s,p}^{(1)}({\bf r'})\varepsilon_1({\bf r'}) \right>\right)
\end{gather}
\end{widetext}
, where ${\bf R}_{p,0}^{spec}$ is the reflection coefficient in case of the absence of the roughness and ${\bf R}_{p,i1}^{spec}$ and ${\bf R}_{p,i2}^{spec}$ are corrections, contributed by the interference terms. Since for $z'$ inside the medium all fields and Green's functions are $\sim e^{ik_{-}z'}$, we get for the first interference term a multiplier $2ik_{-}$:
\begin{widetext}
\begin{eqnarray}
    {\bf R}_{p,i1}^{spec}=-4k_0\frac{\delta^2}{a} Re\left[ r_p^* t_p\frac{(\varepsilon-1)}{\cos\theta} \sqrt{\varepsilon-\sin^2\theta} \left( G_{xx}^{(+-)}(r=\beta\sin\theta)\sqrt{1-\frac{\sin^2\theta}{\varepsilon}} + G_{xz}^{(+-)}(r=\beta\sin\theta) \frac{\sin\theta}{\sqrt{\varepsilon}} \right) \right]
\end{eqnarray}    
\end{widetext}

The second interference term will not contain any derivatives over $z'$ and will result in a form:

\begin{widetext}
    \begin{gather}
    {\bf R}_{p,i2}^{spec}=2\left(\frac{\delta}{a}\right)^2 Im\left[ r_p^* t_p\frac{(\varepsilon-1)^2}{\cos\theta} (G_{xx}^{(+-)}, 0, G_{xz}^{(+-)})|_{r=\beta\sin\theta} \cdot \hat{I}^{(2D)}(\beta,\theta)\cdot \left(\begin{array}{c}
        \sqrt{1-\sin^2\theta / \varepsilon} \\
        0 \\
        \sin\theta / \sqrt{\varepsilon}
    \end{array}\right) \right]
\end{gather}
\end{widetext}

, where we recall the tensor
\begin{widetext}
\begin{gather}
\hat{I}^{(2D)}=\left(\begin{array}{ccc}
I_{xx}^{(2D)} & 0 & I_{xz}^{(2D)}\\
0 & I_{yy}^{(2D)} & 0\\
I_{zx}^{(2D)} & 0 & I_{zz}^{(2D)} \end{array}\right)
\end{gather}
\end{widetext}

with components that are presented in Eq.(\ref{Ixxxzzxzz}).

Finally, writing the total expression for the specular part of the integral scattering coefficient:


\begin{widetext}
\begin{gather}
    \left<{\bf R}_{p}^{spec}\right>=|r_p|^2 + 2\left(\frac{\delta}{a}\right)^2 Im\left[ r_p^* t_p\frac{(\varepsilon-1)^2}{\cos\theta} (G_{xx}^{(+-)}, 0, G_{xz}^{(+-)})|_{r=\beta\sin\theta} \cdot \hat{I}^{(2D)}(\beta,\theta)\cdot \left(\begin{array}{c}
        \sqrt{1-\sin^2\theta / \varepsilon} \\
        0 \\
        \sin\theta / \sqrt{\varepsilon}
    \end{array}\right) \right]-\nonumber\\
    -4k_0\frac{\delta^2}{a} Re\left[ r_p^* t_p\frac{(\varepsilon-1)}{\cos\theta} \sqrt{\varepsilon-\sin^2\theta} \left( G_{xx}^{(+-)}(r=\beta\sin\theta)\sqrt{1-\frac{\sin^2\theta}{\varepsilon}} + G_{xz}^{(+-)}(r=\beta\sin\theta) \frac{\sin\theta}{\sqrt{\varepsilon}} \right) \right]
    \label{RppSpectotal}
\end{gather}
\end{widetext}



Analogously, for incident s-polarization we get for the specular part
\begin{widetext}
    \begin{gather}
    \left<{\bf R}_{s}^{spec}\right>={\bf R}_{s,0}^{spec}+{\bf R}_{s,i1}^{spec}+{\bf R}_{s,i2}^{spec}\nonumber\\
    {\bf R}_{s,0}^{spec}=|r_s|^2\nonumber\\
    {\bf R}_{s,i1}^{spec}=-4k_0\frac{\delta^2}{a}Re\left[r_s^* t_s (\varepsilon-1)\sqrt{\varepsilon-\sin^2\theta}G_{yy}^{(+-)}(r=\beta\sin\theta)\right]\nonumber\\
    {\bf R}_{s,i2}^{spec}=2\left(\frac{\delta}{a}\right)^2Im\left[ r_s^* t_s G_{yy}^{(+-)}(r=\beta\sin\theta) I^{(2D)}_{yy}(\beta,\theta) \right]
\end{gather}
\end{widetext}

and for the diffuse part the differential reflection coefficients are
\begin{widetext}
\begin{gather}
    \left<\left(\frac{\partial R_{ps}}{\partial \Omega_s}\right)_{diff}\right>=\left<\left(\frac{\partial R_{ps}}{\partial \Omega_s}\right)\right>= \frac{2\pi\delta^2}{\lambda^2} \frac{|\varepsilon-1|^2}{\cos\theta} |t_s|^2 e^{-\frac{1}{2}\beta^2(\sin^2\theta+\sin^2\theta_s)} e^{\beta^2\sin\theta\sin\theta_s\cos\phi_s} |G_{xx}^{(+-)}(r=\beta\sin\theta_s)|^2 \sin^2\phi_s\nonumber\\
    \left<\left(\frac{\partial R_{ss}}{\partial \Omega_s}\right)_{diff}\right>= \frac{2\pi\delta^2}{\lambda^2} \frac{\cos^2\theta_s}{\cos\theta}|\varepsilon-1|^2 |t_s|^2 e^{-\frac{1}{2}\beta^2(\sin^2\theta+\sin^2\theta_s)} e^{\beta^2\sin\theta\sin\theta_s\cos\phi_s} |G_{yy}^{(+-)}(r=\beta\sin\theta_s)|^2 \cos^2\phi_s
    \label{diffRdiffusepsss}
\end{gather}
\end{widetext}
which leads to the following expressions for the integral coefficients:
\begin{widetext}
    \begin{gather}
    \left<{\bf R}_{ps}^{diff}\right>=\left<{\bf R}_{ps}\right>=\frac{|\varepsilon-1|^2}{\cos\theta}\left(\frac{\delta}{a}\right)^2|t_s|^2e^{-\frac{1}{2}\beta^2\sin^2\theta}\int_0^{\beta} dr r \frac{\beta}{\sqrt{\beta^2-r^2}}e^{-\frac{1}{2}r^2}|G_{xx}^{(+-)}(r)|^2\frac{I_0-I_2}{2}\Bigg|_{x=r\beta\sin\theta}\nonumber\\
    \left<{\bf R}_{ss}^{diff}\right>=\frac{|\varepsilon-1|^2}{\cos\theta}\left(\frac{\delta}{a}\right)^2|t_s|^2e^{-\frac{1}{2}\beta^2\sin^2\theta}\int_0^{\beta} dr r \frac{\sqrt{\beta^2-r^2}}{\beta}e^{-\frac{1}{2}r^2}|G_{yy}^{(+-)}(r)|^2\frac{I_0+I_2}{2}\Bigg|_{x=r\beta\sin\theta}
\end{gather}
\end{widetext}

\subsubsection{1.1.3 Equivalence between specular/diffuse divisions and coherent/incoherent divisions}

Finally, we refer to an important property common to all quantities derived above, crucial for the comparative analysis between the current theory and the older one. We are going to prove that both division methods to specular and diffuse parts and to coherent and incoherent parts are equivalent. The elements of a scattering matrix can be written in a general form as

\begin{eqnarray}
    R_{\alpha \beta}= r_{\alpha} \delta_{\alpha \beta} (2 \pi)^2 \delta(\boldsymbol{q}-\textbf{e}_x k_x) + Q_{\alpha \beta}
    \label{Rgeneral}
\end{eqnarray}

, where we omitted the explicit $\boldsymbol{q}$-dependence of $Q_{\alpha \beta}$ for simplicity. From this, we can distinguish between the specular and diffuse parts by considering the square of its absolute value:

\begin{gather}
    |R_{\alpha \beta}|^2= \left[|r_{\alpha}|^2 S_0 + 2 Re(r_{\alpha}^* Q_{\alpha \beta}) \right] \delta_{\alpha \beta} (2 \pi)^2 \delta(\boldsymbol{q}-\textbf{e}_x k_x) +\nonumber\\+ |Q_{\alpha \beta}|^2,
\end{gather}

where we used $S_0=(2 \pi)^2 \delta(\boldsymbol{0})$. Since $r_{\alpha}$ is roughness-independent, the averaging over the ensemble gives

\begin{gather}
    \left<|R_{\alpha \beta}|^2\right>= \left[|r_{\alpha}|^2 S_0 + 2 Re(r_{\alpha}^* \left<Q_{\alpha \beta}\right>) \right] \delta_{\alpha \beta} (2 \pi)^2 \delta(\boldsymbol{q}-\textbf{e}_x k_x) + \nonumber\\+\left<|Q_{\alpha \beta}|^2\right>
\end{gather}

On the other hand, if we first average both sides of Eq. (\ref{Rgeneral}) and then calculate the square of the absolute value of the result, we obtain

\begin{gather}
    |\left<R_{\alpha \beta}\right> |^2=\nonumber\\= \left[|r_{\alpha}|^2 S_0 + 2 Re(r_{\alpha}^* \left<Q_{\alpha \beta}\right>) \right] \delta_{\alpha \beta} (2 \pi)^2 \delta(\boldsymbol{q}-\textbf{e}_x k_x).\nonumber\\
\end{gather}

Here, we neglected all terms \(O(\delta^4)\) arising from \(|\left<Q_{\alpha \beta}\right>|^2\) due to the problem's setup. This demonstrates that for all reflection coefficients, the coherent part is simultaneously its specular part, and consequently, the incoherent part is its diffuse part.

\subsection{1.2 Approximation $\beta\gg 1$}

We start with the results of the previous sections:

\begin{widetext}
\begin{gather}
    \left<{\bf R}_{p}^{spec}\right>=|r_p|^2 + 2\left(\frac{\delta}{a}\right)^2 Im\left[ r_p^* t_p\frac{(\varepsilon-1)^2}{\cos\theta} (G_{xx}^{(+-)}, 0, G_{xz}^{(+-)})|_{r=\beta\sin\theta} \cdot \hat{I}^{(2D)}(\beta,\theta)\cdot \left(\begin{array}{c}
        \sqrt{1-\sin^2\theta / \varepsilon} \\
        0 \\
        \sin\theta / \sqrt{\varepsilon}
    \end{array}\right) \right]-\nonumber\\
    -4k_0\frac{\delta^2}{a} Re\left[ r_p^* t_p\frac{(\varepsilon-1)}{\cos\theta} \sqrt{\varepsilon-\sin^2\theta} \left( G_{xx}^{(+-)}(r=\beta\sin\theta)\sqrt{1-\frac{\sin^2\theta}{\varepsilon}} + G_{xz}^{(+-)}(r=\beta\sin\theta) \frac{\sin\theta}{\sqrt{\varepsilon}} \right) \right],\nonumber\\
    \left<{\bf R}_{s}^{spec}\right>=|r_s|^2-4k_0\frac{\delta^2}{a}Re\left[r_s^* t_s (\varepsilon-1)\sqrt{\varepsilon-\sin^2\theta}G_{yy}^{(+-)}(r=\beta\sin\theta)\right]+2\left(\frac{\delta}{a}\right)^2Im\left[ r_s^* t_s G_{yy}^{(+-)}(r=\beta\sin\theta) I^{(2D)}_{yy}(\beta,\theta) \right],\nonumber\\
    \left<{\bf R}_{pp}^{diff}\right>=\frac{|\varepsilon-1|^2}{\cos\theta} \left(\frac{\delta}{a}\right)^2 |t_p|^2 e^{-\frac{1}{2}\beta^2\sin^2\theta} \int_0^{\beta}dr r \frac{\beta}{\sqrt{\beta^2-r^2}} e^{-\frac{1}{2} r^2}\times \nonumber\\ \times \left[ |G_{xx}^{(+-)}(r)|^2\cdot \left| 1-\frac{\sin^2\theta}{\varepsilon} \right|\cdot  \frac{I_0+I_2}{2}\Bigg|_{x=r\beta\sin\theta}+2 Re\left(\left[ G_{xx}^{(+-)} (r)\sqrt{ 1-\frac{\sin^2\theta}{\varepsilon} }  \text{ } \right]^{*}G_{xz}^{(+-)}(r)\frac{\sin\theta}{\sqrt{\varepsilon}}\right)I_1\Bigg|_{x=r\beta\sin\theta} +\right.\nonumber\\ \left.+ |G_{xz}^{(+-)}(r)|^2\frac{\sin^2\theta}{|\varepsilon|} I_0\Bigg|_{x=r\beta\sin\theta} \right], \nonumber\\
    \left<{\bf R}_{sp}^{diff}\right>=\left<{\bf R}_{sp}\right>=\frac{|\varepsilon-1|^2}{\cos\theta} \left(\frac{\delta}{a}\right)^2 |t_p|^2 e^{-\frac{1}{2}\beta^2\sin^2\theta}  \int_0^{\beta}dr r \frac{\sqrt{\beta^2-r^2}}{\beta} e^{-\frac{1}{2} r^2} |G_{yy}^{(+-)}(r)|^2\frac{I_0-I_2}{2}\Bigg|_{x=r\beta\sin\theta},\nonumber\\
    \left<{\bf R}_{ps}^{diff}\right>=\left<{\bf R}_{ps}\right>=\frac{|\varepsilon-1|^2}{\cos\theta}\left(\frac{\delta}{a}\right)^2|t_s|^2e^{-\frac{1}{2}\beta^2\sin^2\theta}\int_0^{\beta} dr r \frac{\beta}{\sqrt{\beta^2-r^2}}e^{-\frac{1}{2}r^2}|G_{xx}^{(+-)}(r)|^2\frac{I_0-I_2}{2}\Bigg|_{x=r\beta\sin\theta},\nonumber\\
    \left<{\bf R}_{ss}^{diff}\right>=\frac{|\varepsilon-1|^2}{\cos\theta}\left(\frac{\delta}{a}\right)^2|t_s|^2e^{-\frac{1}{2}\beta^2\sin^2\theta}\int_0^{\beta} dr r \frac{\sqrt{\beta^2-r^2}}{\beta}e^{-\frac{1}{2}r^2}|G_{yy}^{(+-)}(r)|^2\frac{I_0+I_2}{2}\Bigg|_{x=r\beta\sin\theta},
    \label{RPandStotal}
\end{gather}
\end{widetext}

where

\begin{eqnarray}
    G_{xx}^{(+-)}(r)=\frac{\sqrt{\varepsilon\beta^2-r^2}\sqrt{\beta^2-r^2}}{\varepsilon \sqrt{\beta^2-r^2}+\sqrt{\varepsilon \beta^2-r^2}},\nonumber\\
    G_{xz}^{(+-)}(r)=-\frac{r \sqrt{\beta^2-r^2}}{\varepsilon \sqrt{\beta^2-r^2}+\sqrt{\varepsilon \beta^2-r^2}},\nonumber\\
    G_{yy}^{(+-)}(r)=\frac{\beta^2}{\sqrt{\beta^2-r^2}+\sqrt{\varepsilon \beta^2-r^2}},
    \label{Gtensor}
\end{eqnarray}

and

\begin{widetext}
\begin{gather}
I_{yy}^{(2D)}(\beta,\theta)=i e^{-\beta^2 \sin^2\theta / 2} \int_{0}^{\infty} dr r  e^{-r^2/2} \times \nonumber\\ \left[ \frac{\sqrt{\beta^2-r^2}\sqrt{\varepsilon \beta^2-r^2}}{\sqrt{\varepsilon \beta^2-r^2}+\varepsilon \sqrt{\beta^2-r^2}} 
\frac{I_0-I_2}{2} \Bigg|_{r\beta\sin\theta}+\frac{\beta^2}{\sqrt{\varepsilon \beta^2-r^2}+\sqrt{\beta^2-r^2}} 
\frac{I_0+I_2}{2} \Bigg|_{r\beta\sin\theta} \right], \nonumber\\
    I^{(2D)}_{xx}(\beta,\theta)= ie^{-\beta^2 \sin^2 \theta/2} \int_0^{\infty}dr r e^{-r^2/2} \times \nonumber \\ \times \left[ \frac{\sqrt{\beta^2-r^2}\sqrt{\varepsilon \beta^2-r^2}}{\sqrt{\varepsilon \beta^2-r^2}+\varepsilon \sqrt{\beta^2-r^2}} \frac{I_0+I_2}{2} \Bigg|_{r\beta\sin\theta} + \frac{\beta^2}{\sqrt{\varepsilon \beta^2-r^2}+\sqrt{\beta^2-r^2}} \frac{I_0-I_2}{2}\Bigg|_{r\beta\sin\theta} \right], \nonumber \\
    I^{(2D)}_{xz}(\beta,\theta)=ie^{-\beta^2 \sin^2 \theta/2} \int_0^{\infty}dr r e^{-r^2/2}
    \left[\frac{r\sqrt{\varepsilon\beta^2-r^2}}{\varepsilon(\sqrt{\varepsilon \beta^2-r^2}+\varepsilon \sqrt{\beta^2-r^2})}I_1(r\beta\sin \theta)\right], \nonumber \\
    I^{(2D)}_{zx}(\beta,\theta)=ie^{-\beta^2 \sin^2 \theta/2} \int_0^{\infty}dr r e^{-r^2/2}
    \left[\frac{r\sqrt{\beta^2-r^2}}{\sqrt{\varepsilon \beta^2-r^2}+\varepsilon \sqrt{\beta^2-r^2}}I_1(r\beta\sin \theta)\right], \nonumber \\
    I^{(2D)}_{zz}(\beta,\theta)=ie^{-\beta^2 \sin^2 \theta/2} \int_0^{\infty}dr r e^{-r^2/2}
    \left[\frac{r^2}{\varepsilon(\sqrt{\varepsilon \beta^2-r^2}+\varepsilon \sqrt{\beta^2-r^2})}I_0(r\beta\sin \theta)\right].
    \label{Ixxxzzxzz}
\end{gather}
\end{widetext}

Generally, these integrals cannot be simplified into an analytical form. However, we proceed by considering the case of normal incidence for \(\beta \gg 1\), which provides a good approximation for experimental purposes.

\subsubsection{1.2.1 Specular scattering}





For a normal incidence only $I^{2D}_{xx}$ and $I^{2D}_{yy}$ components of the tensor $\hat{I}^{2D}$ become important and the argument $r\beta\sin \theta$ vanishes for any value of $r$. Hence, the modified Bessel functions reduce to a constant value:

    \begin{eqnarray}
        I_{xx}^{(2D)}(\beta,0)=I_{yy}^{(2D)}(\beta,0)=\frac{1}{2}i  \int_{0}^{\infty} dr r  e^{-r^2/2} \times \nonumber\\ \left[ \frac{\sqrt{\beta^2-r^2}\sqrt{\varepsilon \beta^2-r^2}}{\sqrt{\varepsilon \beta^2-r^2}+\varepsilon \sqrt{\beta^2-r^2}} 
+\frac{\beta^2}{\sqrt{\varepsilon \beta^2-r^2}+\sqrt{\beta^2-r^2}} 
 \right]\nonumber\\
    \end{eqnarray}

Because of the exponential only the values of $r\lesssim 1$ give reasonable contribution to the integral, for which the expression in the square brackets reduces to a constant because of $\beta \gg 1$. With this approximation we obtain:

\begin{eqnarray}
    I_{xx}^{(2D)}(\beta,0)=I_{yy}^{(2D)}(\beta,0)\approx \frac{i\beta}{1+\sqrt{\varepsilon}}.
\end{eqnarray}

With this for the specular reflection coefficients we have the asymptotics:

\begin{gather}
    {\bf R}_{p,i1}^{spec}=-8k_0^2\delta^2 \left|\frac{\sqrt{\varepsilon}-1}{\sqrt{\varepsilon}+1}\right|^2 n,\nonumber\\
    {\bf R}_{p,i2}^{spec}=4k_0^2\delta^2 \left|\frac{\sqrt{\varepsilon}-1}{\sqrt{\varepsilon}+1}\right|^2 (n-1),
\end{gather}

where $n=Re[\sqrt{\varepsilon}]$.

Finally, we can compare our current theoretical results with respect to the old theory by Navarrete \textit{et al.}:

\begin{gather}
    \left<{\bf R}^{spec}_{Old}\right>=\left|\frac{\sqrt{\varepsilon}-1}{\sqrt{\varepsilon}+1}\right|^2 [1-4k_0^2\delta^2],\nonumber\\
    \left<{\bf R}^{spec}_{Current}\right>=\left|\frac{\sqrt{\varepsilon}-1}{\sqrt{\varepsilon}+1}\right|^2 [1-4k_0^2\delta^2 (n+1)].
\end{gather}

Because of the presence of $n$ the influence of the correction terms becomes much higher and the anti-reflection effect becomes more expressive.

\subsubsection{1.2.2 Diffuse scattering}

Performing considerations described above we can obtain the results for the diffuse part:

\begin{eqnarray}
    \left<{\bf R}^{diff}_{Old}\right>=k_0^2\delta^2\left|\frac{\sqrt{\varepsilon}-1}{\sqrt{\varepsilon}+1}\right|^2,\nonumber\\
    \left<{\bf R}^{diff}_{Current}\right>=2k_0^2\delta^2\left|1-\frac{1}{\sqrt{\varepsilon}}\right|^2
\end{eqnarray}

where we omitted indices of polarizations for the scattered ($\alpha$) and the incident ($\beta$) beams.

\subsection{1.3 Approximation $\beta\ll 1$}

Here we proceed by considering the case of \(\beta \ll 1\), which provides a good approximation for typical values of reflection coefficients, for example, in the industry of microelectronics.

\subsubsection{1.3.1 Specular scattering}

Here we will consider only the case of the normal incidence of light, when Fresnel coefficients become:

\begin{eqnarray}
        r = \frac{\sqrt{\varepsilon}-1}{\sqrt{\varepsilon}+1},\nonumber\\
        t = \frac{2}{\sqrt{\varepsilon}+1}.
\end{eqnarray}

and for tensor $\hat{G}$ from Eq. (\ref{Gtensor}) we have:

\begin{gather}
    G_{xx}^{(+-)}(r=0)=G_{yy}^{(+-)}(r=0)=\frac{\beta}{1+\sqrt{\varepsilon}},\nonumber\\
    G_{xz}^{(+-)}(r=0)=0.\nonumber\\
\end{gather}

Also, we put $\beta=0$ in Eq. (\ref{Ixxxzzxzz}) to obtain the leading order terms:

\begin{gather}
    I^{(2D)}_{xx}=I^{(2D)}_{yy}=-\frac{\sqrt{2\pi}}{4}\frac{1}{\varepsilon+1},
\end{gather}

since we assume that even stronger confinement $\sqrt{|\varepsilon|}\beta\ll 1$ holds. The analytical result, obtained in this way:

\begin{eqnarray}
    \left<R^{spec}\right>=\left| \frac{\sqrt{\varepsilon}-1}{\sqrt{\varepsilon}+1} \right|^2\cdot \left[1 - 4\sqrt{2\pi}\frac{n}{|\varepsilon+1|^2}\frac{\delta^2}{ad}-8k_0^2\delta^2 n\right].\nonumber\\
\end{eqnarray}

contains an additional parameter, that is characteristic to an opaque system: the skin depth $d=(\kappa k_0)^{-1}$, where $\kappa=Im(\sqrt{\varepsilon})$.

\subsubsection{1.3.2 Diffuse scattering}

Consider the differential coefficients for diffuse reflection. We need to introduce small and large angles in a following way: $\sqrt{|\varepsilon|}\cos\theta \gg 1$ represent small angles and $\sqrt{|\varepsilon|}\cos\theta \ll 1$ represent large angles. Then, we consider the limit $|\varepsilon|\gg 1$.

With this we can write the following approximations: For small angles $\theta$ and $\theta_s$

\begin{gather}
    G_{xx}^{(+-)}(r=\beta\sin\theta_s)=\frac{\beta}{\sqrt{\varepsilon}},\nonumber\\
    G_{xz}^{(+-)}(r=\beta\sin\theta_s)=-\frac{\beta}{\varepsilon}\sin\theta_s,\nonumber\\
    t_p (\theta)= \frac{2}{\sqrt{\varepsilon}},\nonumber\\
\end{gather}

while for the large ones

\begin{gather}
    G_{xx}^{(+-)}(r=\beta\sin\theta_s)=\beta\cos\theta_s,\nonumber\\
    G_{xz}^{(+-)}(r=\beta\sin\theta_s)=-\frac{\beta}{\sqrt{\varepsilon}}\sin\theta_s\cos\theta_s,\nonumber\\
    t_p (\theta)= 2\cos\theta.\nonumber\\
\end{gather}

The asymptotic expression $G_{yy}^{(+-)}(r=\beta\sin\theta_s)=\frac{\beta}{\sqrt{\varepsilon}}$ is independent of angle.

Since we have $|G_{xx}^{(+-)}(\beta\sin\theta_s)| \gg |G_{xz}^{(+-)}(\beta\sin\theta_s)|$ for any angle $\theta_s$, we can neglect the influence of the latter and the expressions for the diffuse reflection coefficients in Eq.(\ref{diffRdiffuseppsp}) and Eq.(\ref{diffRdiffusepsss}) reduce to

\begin{gather}
    \left<\left(\frac{\partial R_{ps}}{\partial \Omega_s}\right)_{diff}\right>= \frac{8\pi\delta^2}{\lambda^2} \beta^2\cos\theta \sin^2\phi_s\nonumber\\
    \left<\left(\frac{\partial R_{ss}}{\partial \Omega_s}\right)_{diff}\right>= \frac{8\pi\delta^2}{\lambda^2} \cos\theta \cos^2\theta_s \cos^2\phi_s\nonumber\\
\end{gather}

while for other two components we have to distinguish between small and large angles:

\begin{gather}
    \left<\left(\frac{\partial R_{pp}}{\partial \Omega_s}\right)_{diff}\right>= \frac{8\pi\delta^2}{\lambda^2} \beta^2 \frac{\cos^2\phi_s}{\cos\theta}\nonumber\\
    \left<\left(\frac{\partial R_{sp}}{\partial \Omega_s}\right)_{diff}\right>= \frac{8\pi\delta^2}{\lambda^2}  \frac{\cos^2\theta_s \sin^2\phi_s}{\cos\theta}\nonumber\\
\end{gather}

for small incidence angles and 

\begin{gather}
    \left<\left(\frac{\partial R_{pp}}{\partial \Omega_s}\right)_{diff}\right>= \frac{8\pi\delta^2}{\lambda^2} \beta^2 |\varepsilon|\cos\theta\cos^2\phi_s\nonumber\\
    \left<\left(\frac{\partial R_{sp}}{\partial \Omega_s}\right)_{diff}\right>= \frac{8\pi\delta^2}{\lambda^2}  |\varepsilon|\cos\theta\cos^2\theta_s \sin^2\phi_s\nonumber\\
\end{gather}

for large incidence angles. Here we kept the scattering angle $\theta_s$ small, since they occupy the most of the available domain. With this we see, that haze intensity $\sim k_0^4\delta^2 a^2$ for $\beta\ll1$.

\section{Note 2: Experiment}

\begin{figure}
\begin{tabular}{c}
    \begin{overpic}[width=0.9\columnwidth,keepaspectratio]{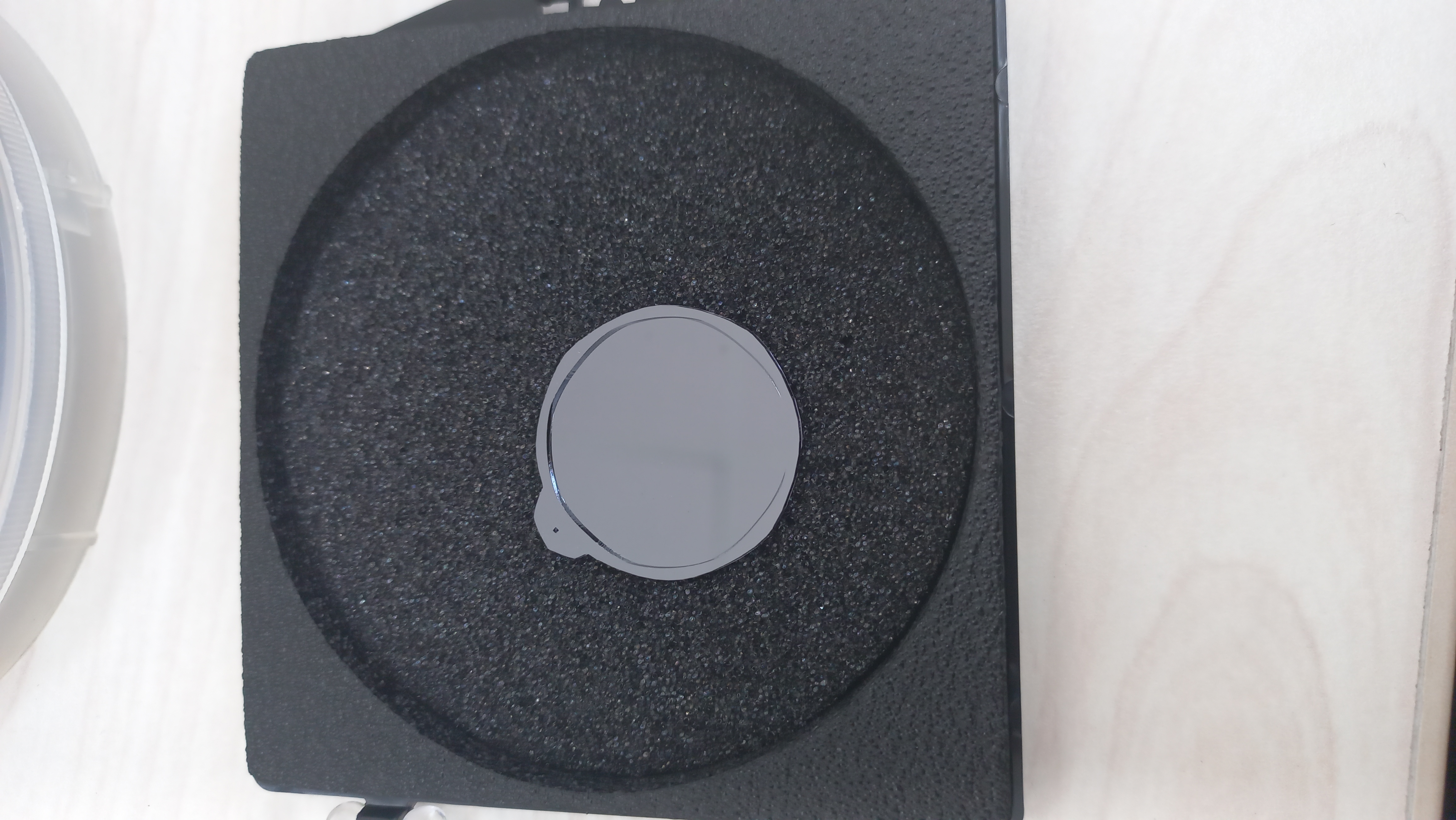}
    \put(0,0.47\columnwidth){(a)}
\end{overpic} \\
    \begin{overpic}[width=0.9\columnwidth,keepaspectratio]{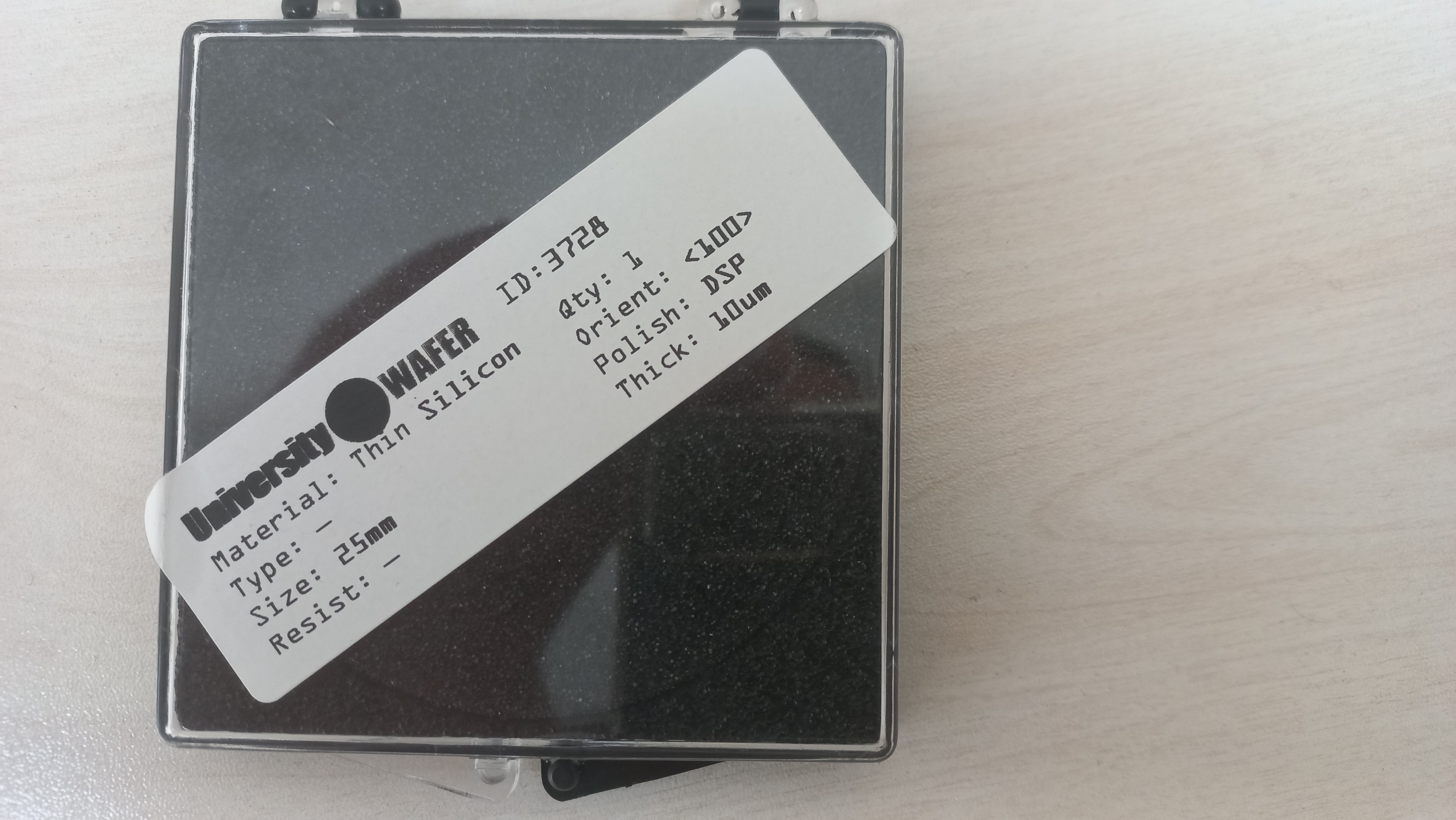}
    \put(0,0.47\columnwidth){(b)}
\end{overpic}\\
\end{tabular}
\caption{Photography of the sample (a) and its characteristics (b).}
\label{samples}
\end{figure}

The 10 \(\mathrm{\mu}\)m thick monocrystalline silicon wafers with well-polished surfaces (University Wafers, ID
3728, orientation <100>, see Fig. \ref{samples}) were first cleaned using a plasma cleaning method to remove any
surface contaminants. The etching solution was prepared by mixing 4.57 vol$\%$ hydrofluoric acid
(HF, 49$\%$), 4.57 vol$\%$ hydrogen peroxide ($\text{H}_2\text{O}_2$, 30$\%$), and 49.10 vol$\%$ hydrochloric acid (HCl,
37$\%$) in deionized water. The silicon wafers were immersed in this mixture for 1 second at room
temperature to achieve a roughness of 8-9 nm. After etching, the wafers were immediately
rinsed with deionized water and dried using nitrogen gas.
Diffuse reflection measurements were performed using an Agilent Cary 60 spectrometer, both
when the wafers were flat and after creating roughness. All traces were corrected for baseline
and scanned within the range of 300-800 nm. A white Polytetrafluoroethylene (PTFE) sample
was used to establish the 100$\%$ reflectance baseline. These measurements provided a
comparative analysis of the diffuse reflection spectra before and after the etching process,
highlighting the changes in surface properties due to the induced roughness.
Atomic Force Microscopy (AFM) measurements were conducted using an LS-AFM atomic force
microscope (AFM Workshop, USA) operating in tapping mode in the air at room temperature.
Silicon cantilever tips ACLA-10-W with reflective Al (back side) coating (Budget Sensors,
Innovative Solutions Ltd, Bulgaria) was used. Subsequently, all the images were flattened by
means of the AFM Control software. The software used for topographic data results
analysis and particle size determination is Gwyddion 2.60. The AFM analysis confirmed that the
surface roughness of the etched silicon wafers was consistently within the desired range of 8-9
nm.

\section{Note 3: Results}

\begin{figure}
    \centering
    \includegraphics[width=\linewidth]{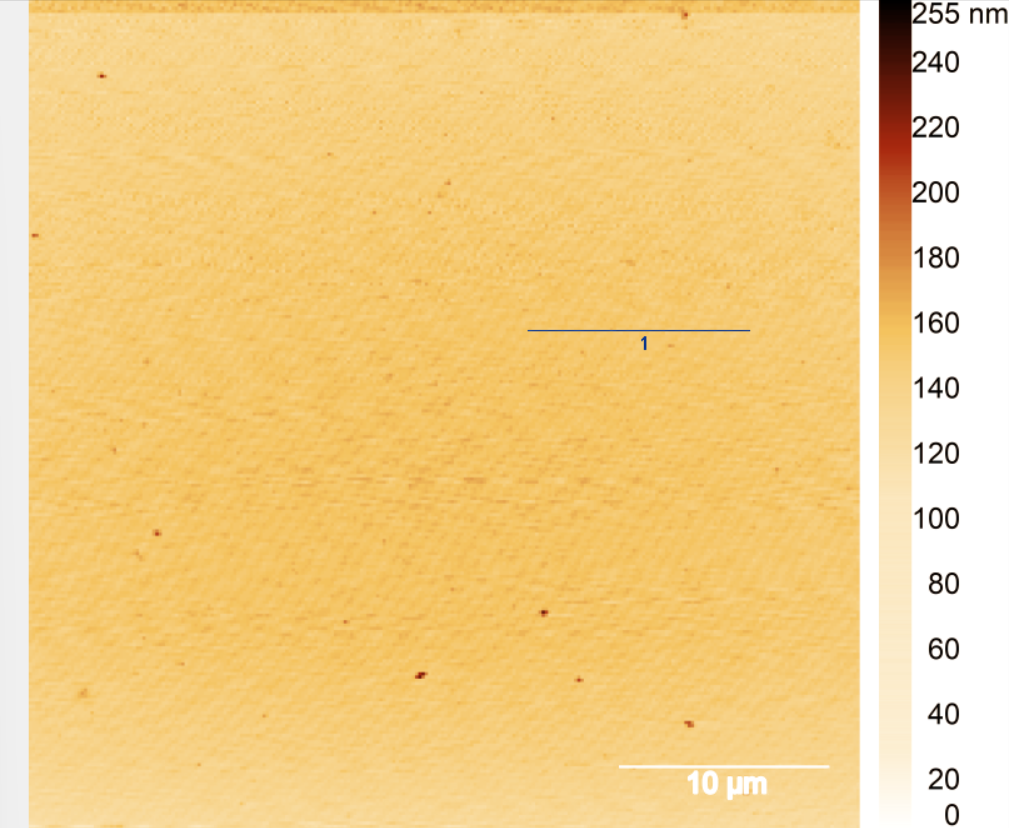}
    \caption{AFM image of the surface of a rough sample. The scale represents the displacement of surface points along the z-direction. Line 1 denotes the path along which the data was acquired.}
    \label{fig:enter-label1}
\end{figure}





Finding the correlation function of a rough surface follows the approach outlined in \cite{bennett1976} as follows: First, we calibrate the shifts so that their mean value is zero. Next, we define the distance between two neighboring points as a unit length, \( \rho_0 = 2.0696 \) nm, and compute the correlation function using the formula

\begin{eqnarray}
    W(\rho) = W(m\rho_0) = \frac{1}{M} \sum_{n=1}^{M-m} z(n\rho_0) z(n\rho_0+m\rho_0),
\end{eqnarray}

where \( M = 4081 \). Since the values of \( \rho \ll (M-1)\rho_0 = 8.44 \) $\mu$m are minimally affected, we focus our analysis on the vicinity of \( \rho = 0 \) and fit the function \( W(x) = \delta^2 e^{-x^2/2a^2} \) to this domain. 


Subsequently, with the obtained fit parameters:

\begin{eqnarray}
    \delta=\text{(}7.92383 \pm 0.0438912\text{) nm} \nonumber\\
    a=\text{(}28.7482 \pm 0.751428\text{) nm}
    \label{fitparams}
\end{eqnarray}

we proceed to simulate the resulting reflection coefficient according to various theories. 

To compare the theory presented in this paper with that in Navarrete \textit{et al.} (2009) \cite{navarrete2009}, we conduct a comparative analysis. 

First, we verify that the data from \cite{green2008} accurately reproduces the scattering results from the rough surface, ensuring consistency with experimental findings (see "Experimental Data Smooth" and "Theory Smooth"). Subsequently, we simulate the reflection coefficient for both theories for a rough surface using parameters as described in Eq. (\ref{fitparams}) (see "Old Theory Rough" and "New Theory Rough"). According to Fig. 3, the predictions of the new theory exhibit a closer alignment with experimental observations (see "Experimental Data Rough").

\subsection{3.1 Additional results}


For the industry of microelectronics typical values of the roughness RMS are $\delta\lesssim 1-2 \text{ nm}$ \cite{mori2020,wang2005,khan2007}. Thus, investigation of the behavior of haze for $\delta=2 \text{ nm}$ satisfies industrial needs. 

\begin{figure}
    \centering
    \begin{tabular}{c}
       \begin{overpic}[width=\columnwidth,keepaspectratio]{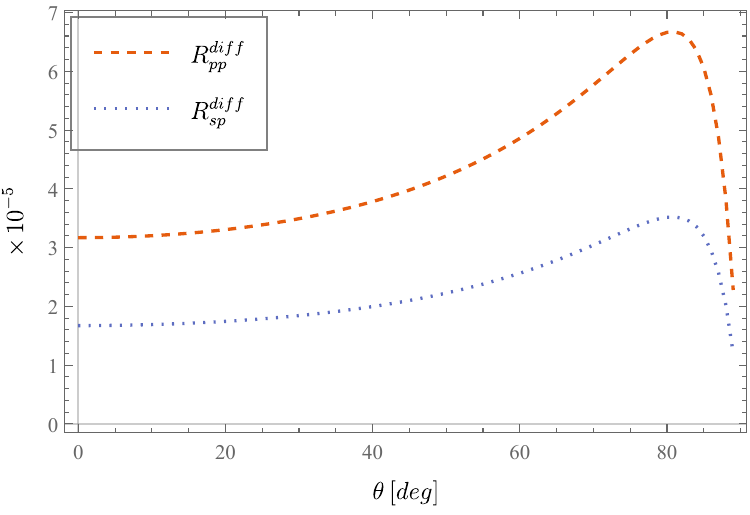}
    \put(0,0.65\columnwidth){(a)}
\end{overpic} \\
    \begin{overpic}[width=\columnwidth,keepaspectratio]{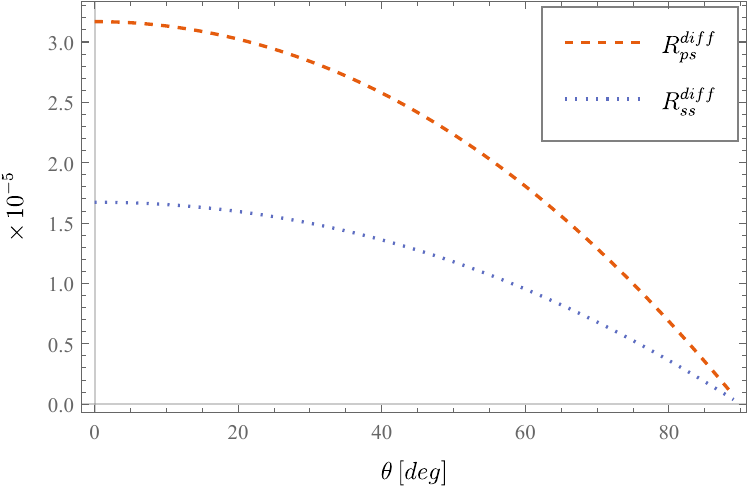}
    \put(0,0.635\columnwidth){(b)}
\end{overpic}\\
    \end{tabular}
    \caption{The angular dependence of the integrated diffuse reflection coefficients for incident p- (a) and s-polarized (b) waves is shown. Simulations were conducted with a roughness RMS $\delta = 2$ nm, correlation length $a = 10$ nm, and wavelength $\lambda = 350$ nm.}
    \label{diffuseintegral}
\end{figure}
\begin{figure*}
\begin{tabular}{cc}
    \begin{overpic}[width=\columnwidth,keepaspectratio]{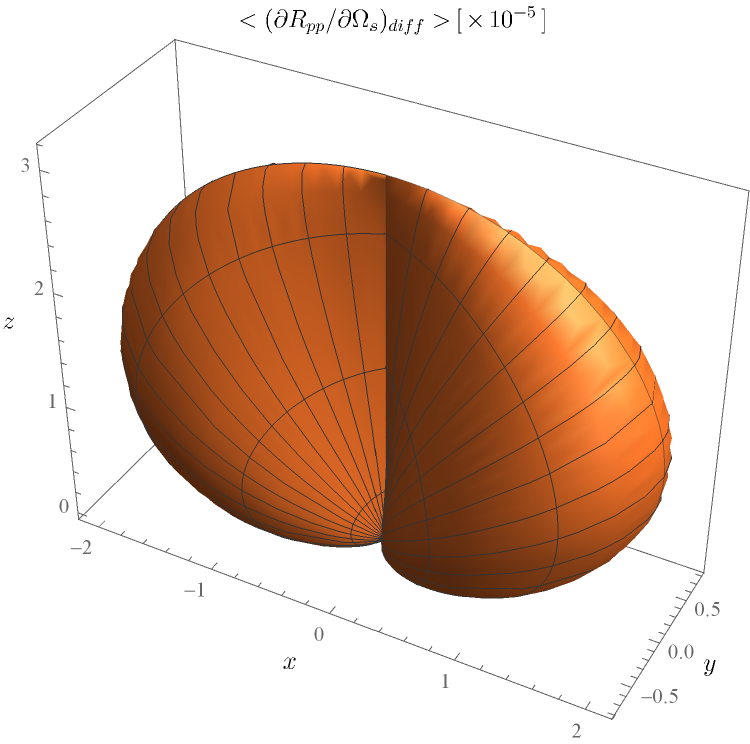}
    \put(0,\columnwidth){(a)}
\end{overpic} &
    \begin{overpic}[width=0.75\columnwidth,keepaspectratio]{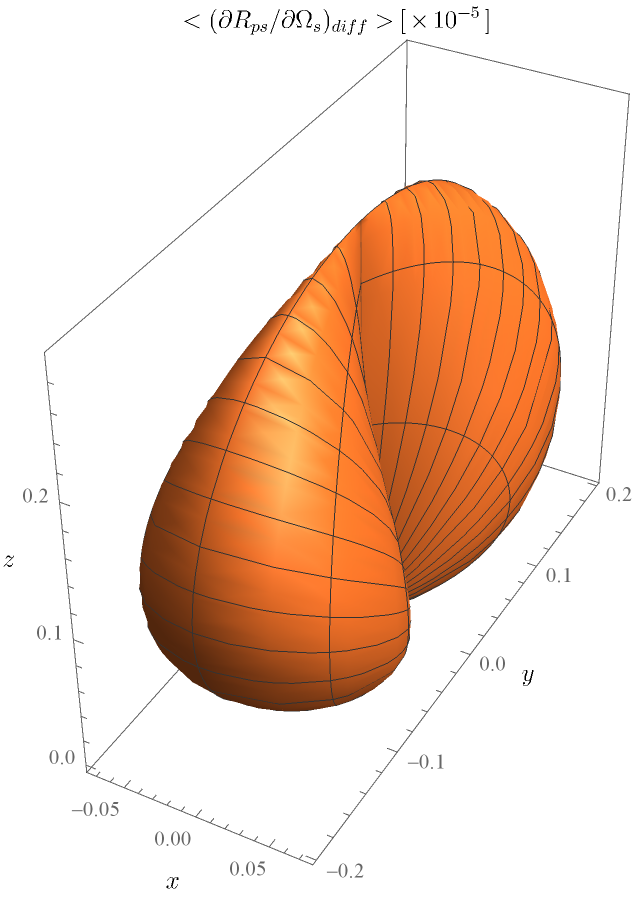}
    \put(0,\columnwidth){(b)}
\end{overpic}\\
    \begin{overpic}[width=0.75\columnwidth,keepaspectratio]{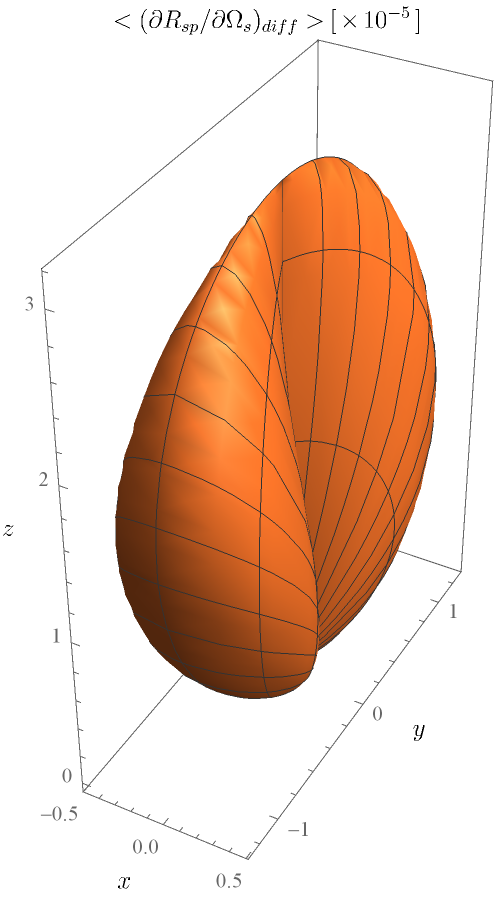}
    \put(0,\columnwidth){(c)}
\end{overpic} &
    \begin{overpic}[width=\columnwidth,keepaspectratio]{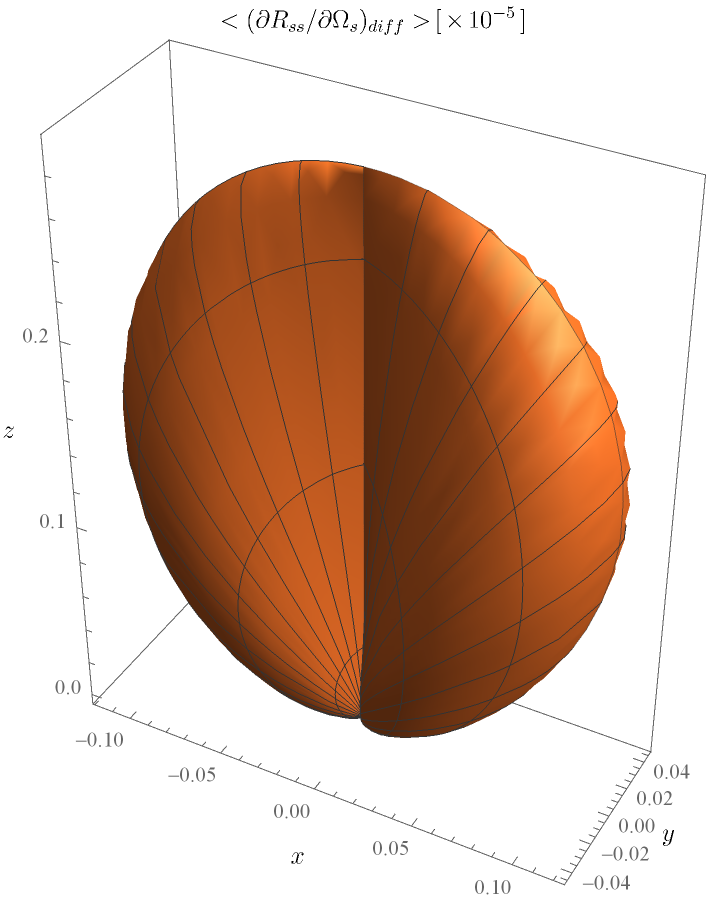}
    \put(0,\columnwidth){(d)}
\end{overpic}\\
\end{tabular}
\caption{The polar plots depict the differential reflection coefficients (in $\text{str}^{-1}$) for diffuse p- (a, b) and s-polarized (c, d) scattering, considering incident p- (a, c) and s-polarized (b, d) waves. Simulations were conducted with an incident angle of $\theta = 80^\circ$, roughness RMS $\delta = 2$ nm, correlation length $a = 10$ nm, and wavelength $\lambda = 350$ nm.}
\label{DiffuseDiffR}
\end{figure*}

Figs. \ref{diffuseintegral}-\ref{DiffuseDiffR} illustrate the primary characteristics of diffuse scattering. The diffusely scattered light is predominantly p-polarized, a behavior that remains independent of the incident polarization, and is directed along the surface normal, without regard to the angle of incidence. Maximum haze originates for incident p-polarization at large angles of incidence, close to Brewster's angle.

\clearpage